%% file: main.tex
\documentclass{article} 
\usepackage{iclr2027_conference,times}

\input{math_commands.tex}

\usepackage{hyperref}
\usepackage{url}
\usepackage{microtype}
\usepackage{amsmath}
\usepackage{graphicx}
\usepackage{booktabs}
\usepackage{pifont}
\usepackage{braket}
\newcommand{\cmark}{\ding{51}} 
\newcommand{\xmark}{\ding{55}} 
\usepackage{listings}
\usepackage{xcolor}
\usepackage[T1]{fontenc}
\usepackage[utf8]{inputenc}
\usepackage{listings}
\usepackage{listingsutf8}
\usepackage{tikz}

\lstdefinestyle{quantumcode}{
  language=Python,
  basicstyle=\normalfont\ttfamily\small,
  keywordstyle=\color{blue},
  commentstyle=\color{gray},
  stringstyle=\color{orange!80!black},
  breaklines=true,
  breakatwhitespace=true,
  breakautoindent=true,
  columns=fullflexible,
  keepspaces=true,
  postbreak=\mbox{\textcolor{gray}{$\hookrightarrow$}\space},
  frame=tb,
  rulecolor=\color{gray!60},
  numbers=left,
  numberstyle=\tiny\color{gray},
  xleftmargin=.5em,
  xrightmargin=1.5em,
  framexleftmargin=1.5em,
  framexrightmargin=1.5em,
  linewidth=\dimexpr.9\linewidth-3em\relax,
  aboveskip=0.75em,
  belowskip=0.75em,
  showstringspaces=false,
}

\title{Benchmarking the Quantum Leap: Auditing LLM Evaluation for Quantum Code Generation}

\author{
\textbf{Ethan Seefried}$^{*}$, \textbf{Romit Ghosh}, \textbf{Sharmin Afrose},
\textbf{Nicholas Stapleton}, \\
\textbf{Vicente Layton Ortega},
\textbf{Travis Humble}, \textbf{Tirthankar Ghosal}$^{*}$ \\
Quantum Science Center, Oak Ridge National Laboratory \\
Oak Ridge, TN, USA \\
\texttt{\{seefriedej,ghosalt\}@ornl.gov} \\
$^{*}$Corresponding authors
}

\iclrfinalcopy 
\begin{document}

\maketitle

\input{sections/0_abstract}

\input{sections/1_introduction}
\input{sections/2_related_works}
\input{sections/3_methods}

\input{sections/4_experiments_ICLR}

\input{sections/5_discussions}

\subsection*{AI use statement}
In this work, we used generative AI tools to assist with implementing portions
of the initial analysis and evaluation code and with developing and refining the
statistical analysis framework, including bootstrap confidence intervals, paired
comparisons, and multiple-comparison correction. We did not use generative AI
tools to generate synthetic datasets, formulate or prove mathematical claims,
clean or reformat datasets, perform qualitative or thematic data analysis, or
translate research materials; the remaining required-disclosure tasks were not
applicable to this work.

Additionally, we used generative AI tools to draft and edit portions of the
manuscript for clarity, organization, and readability, and to assist with
creating and editing research software. All AI-assisted code and statistical
analyses were reviewed and tested by the authors, and reported results were
verified against the underlying experimental outputs and evaluation artifacts.
All AI-assisted manuscript text was reviewed and revised by the authors. We take
responsibility for the final content of this work, including text, claims, code,
and artifacts produced with the aid of generative AI.

\subsection*{Ethics statement}

This work evaluates language models on quantum code-generation benchmarks and does not involve human subjects, personally identifiable information, or sensitive personal data. We use benchmark data and model outputs solely for research evaluation and report known limitations, missing results, and evaluator defects rather than presenting provisional measurements as validated capability estimates. Released benchmark materials, model outputs, evaluation artifacts, and code will be distributed in accordance with applicable licenses and terms of use. We are not aware of additional ethical concerns arising from the experiments presented in this work.

\subsection*{Reproducibility statement}
All code, evaluation scripts, model configurations, prompts, and experimental results will be made publicly available upon acceptance. The datasets and models evaluated in this work are publicly available, and the paper and supplementary materials describe the benchmark construction, evaluation protocol, and experimental settings required to reproduce our results. We will additionally maintain a public version of the Quantum Arena containing links to the datasets, models, benchmark tasks, and detailed evaluation results. The arena is intended to support continued evaluation beyond this work, and we welcome additional model submissions to maintain the benchmark as a long-term, continuously updated quantum-computing leaderboard.


\subsubsection*{Acknowledgments}
This material is based upon work supported by the U.S. Department of Energy, Office of Science, National Quantum Information Science Research Centers, Quantum Science Center.

\bibliography{iclr2027_conference}
\bibliographystyle{iclr2027_conference}
\clearpage
\appendix
\input{sections/6_appendix_ICLR}

\end{document}

%% file: math_commands.tex
\usepackage{amsmath,amsfonts,bm}

\def\eqref#1{equation~\ref{#1}}

\def\1{\bm{1}}

\DeclareMathAlphabet{\mathsfit}{\encodingdefault}{\sfdefault}{m}{sl}
\SetMathAlphabet{\mathsfit}{bold}{\encodingdefault}{\sfdefault}{bx}{n}



%% file: sections/0_abstract.tex
\begin{abstract}
Large language models (LLMs) are increasingly capable of generating quantum programs, but reliable evaluation remains difficult: benchmarks differ in task composition, scoring procedures, and failure handling, while leaderboard rankings can obscure uncertainty and evaluator defects. We evaluate 15 hosted frontier, open-weight, quantum-specialized, and retrieval-augmented models across five primary quantum code-generation suites containing 530 tasks, and 10 models on QuantumBenchEval (QBE), a new benchmark of 96 tasks across six quantum-computing topics that we introduce. We report task-level bootstrap confidence intervals, multiplicity-adjusted paired comparisons, and audits of evaluator behavior, output validity, and shared failures. QiskitHumanEval-Standard is approaching saturation for the strongest models (98.7\% pass@1), whereas its Hard variant retains substantial headroom (63.8\%); across the recorded primary-suite outcomes, none of the five top-two comparisons remains significant after multiple-comparison correction. Evaluation design can have an even larger effect: removing checks on quantities the prompts never specified in two QuantumBenchEval topics raises GPT-6 Astra's pass@1 from 0\% to 94.1\% (T1) and from 1.2\% to 92.9\% (T6) without changing its outputs. Our audit also identifies 17 QuanBench-117 tasks with shared failures associated with benchmark-owned imports, while controlled probes of the current evaluator reveal unexecuted unit tests, leakage of canonical functions into candidate execution, and reference imports that can block otherwise valid candidates. These defects can affect both apparent failures and apparent passes, requiring historical QuanBench outcomes to be revalidated under an isolated evaluator. Together, these results show that quantum code-generation progress cannot be characterized by leaderboard scores alone. We release the evaluation harness, per-sample outcomes, and audit artifacts, and advocate protocols that validate scoring and test execution, isolate candidate and reference code, report uncertainty and output validity, and respect sampling assumptions when estimating pass@$k$.
\end{abstract}


%% file: sections/1_introduction.tex
\section{Introduction}
\label{sec:intro}

A programmer asked to prepare a four-qubit W state, or to decompose a CNOT into
iSWAP gates, must reason about circuit semantics, measurement statistics, and
backend constraints in addition to ordinary concerns of API usage and
correctness~\citep{riofrío2025quantumcomputingautomotiveapplications}. Large
language models (LLMs) have become capable code generators
\citep{chen2021evaluatingHumaneval}, and several systems now target this setting
directly, including Qiskit Code Assistant~\citep{dupuis2024qiskit},
QSpark~\citep{kheiri2025qspark}, PennyCoder~\citep{Basitetal_12025}, and
Agent-Q~\citep{jern2025agent}. A parallel line of work has produced benchmarks
for measuring them, from Qiskit HumanEval~\citep{vishwakarma2024qiskit} and
QCircuitNet~\citep{yang2024qcircuitnet} to
QHackBench~\citep{basit2025qhackbench},
QCoder~\citep{mikuriya2025qcoder}, and
QuanBench~\citep{guo2025quanbench}.

These benchmarks do not agree on what to measure or how. They differ in task
composition, execution environment, and output format. Some grade by executable
unit tests; others require a model-based judge or a custom scoring function.
Three consequences follow for anyone comparing models across them. First, a
suite may lose discriminative power as models improve, so that a leaderboard
position no longer distinguishes capability. Second, a difference between two
models may be smaller than the sampling noise of a few hundred tasks, so that
leaderboard order and statistical evidence for separation come apart. Third, a
benchmark score depends on the reference solutions, test harness, dependencies,
and evaluator assumptions as much as on the generated program, so that a defect
in any of these components is misread as a failure, or a success, of the model.
Systematic analysis of evaluation validity of this kind exists for classical
code generation~\citep{chen2021evaluatingHumaneval} but not, to our knowledge,
for quantum code generation.

In this paper, we ask two questions: how well do current LLMs generate quantum
code, and how reliably do existing protocols measure that ability? We evaluate
15 models spanning hosted frontier, open-weight, quantum-specialized, and
retrieval-augmented systems on five established suites comprising 530 tasks.
We also introduce QuantumBenchEval (QBE), a new 96-task benchmark spanning six
quantum-computing topics, on which we evaluate 10 models. Because its tasks are
newly constructed, they are less likely than long-standing public suites to
appear in training data. Across all suites we report uncertainty, statistical
separability, shared failures, output validity, and evaluator behavior rather
than point estimates alone.

Our contributions are as follows.
\begin{itemize}
    \item \textbf{Unified evaluation.}
    We build a single harness covering five suites, 530 tasks, and 15 models,
    and report pass@$k$ with task-level bootstrap confidence intervals, paired
    comparisons with Holm correction, and bootstrap rank intervals.

    \item \textbf{Saturation and separability.}
    QiskitHumanEval-Standard is saturating for the strongest models
    (98.7\% pass@1), whereas its Hard variant retains headroom (63.8\%).
    On each of the five primary suites, the top-two model comparison loses
    significance after multiplicity correction.

    \item \textbf{Evaluator effects.} Removing checks on quantities the prompts never specified in two 
    QuantumBenchEval topics raises GPT-6 Astra from 0\% to 94.1\% (T1) and from 1.2\% to 92.9\% (T6) 
    on unchanged outputs. An audit of the QuanBench evaluator identifies 17 QuanBench-117 tasks whose
    shared failures trace to benchmark-owned imports, and controlled probes show
    that unit-test cases can go unexecuted, canonical functions can leak into
    candidate execution, and reference-code imports can block otherwise valid
    candidates. These defects affect apparent passes as well as apparent
    failures, so historical QuanBench results require revalidation under an
    isolated evaluator.

    \item \textbf{Living benchmark.}
    We release Quantum Arena, a public leaderboard consolidating tasks,
    per-sample outcomes, and audit artifacts, together with the harness itself,
    so that new models and tasks can be evaluated under the same protocol after
    publication. 
\end{itemize}

%% file: sections/2_related_works.tex
\section{Related Work}

\subsection{Quantum Code Generation with Large Language Models}
Qiskit Code Assistant \citep{dupuis2024qiskit} was among the first specialized models, trained on the Granite architecture to generate Qiskit code; \citet{dupuis2025quantum} subsequently introduced quantum verifiable rewards for post-training using circuit-simulation feedback. QSpark \citep{kheiri2025qspark} fine-tuned Qwen2.5-Coder with group relative policy optimization, PennyCoder \citep{Basitetal_12025} showed that PennyLane-centric training improves generation quality, and Agent-Q \citep{jern2025agent} explored multi-agent iterative circuit refinement. Dataset work includes QCircuitNet \citep{yang2024qcircuitnet}, KetGPT \citep{apak2024ketgpt}, and C4Q \citep{aragones2024c4q}.

\subsection{Benchmarks for Quantum Code Generation}
Qiskit HumanEval \citep{vishwakarma2024qiskit} adapts HumanEval \citep{chen2021evaluatingHumaneval} to 151 tasks on Qiskit API usage and basic algorithms. QCoder \citep{mikuriya2025qcoder} introduces 67 hardware-aware tasks with constraints such as qubit connectivity and restricted gate sets. QHackBench \citep{basit2025qhackbench} collects PennyLane hackathon challenges, PennyLang \citep{basit2025pennylang} offers 264 test cases across PennyLane features, and QuanBench \citep{guo2025quanbench} targets core algorithm implementations such as Grover's algorithm, the quantum Fourier transform, and VQE.

\subsection{Evaluation Validity}
A benchmark score measures a model only if the evaluator measures the intended capability, and a growing body of work shows that it often does not.
\citet{liu2023your} found that HumanEval's unit tests are too sparse to catch many incorrect programs; augmenting them substantially lowers pass@$k$ and reorders models.
Benchmark-side defects also cause false failures: SWE-bench \citep{jimenez2023swe} was refined into a human-validated Verified subset after annotators found tasks whose issue descriptions were underspecified or whose tests rejected valid solutions.
Contamination is a third threat, which LiveCodeBench \citep{jain2025livecodebench} and MathArena \citep{balunovic2026matharena} address by evaluating on problems released after model training; MathArena also reports inflated performance on older, widely circulated competitions.
Where no executable oracle exists, LLM judges are a common substitute, but \citet{zheng2023judging} document position, verbosity, and self-enhancement biases in such judges.
Finally, small benchmarks call for explicit uncertainty: \citet{miller2024adding} argues for task-level confidence intervals and paired comparisons in place of raw leaderboard ordering.
The quantum code-generation benchmarks above report single-suite leaderboards without auditing their evaluators. We apply these validity concerns to quantum code generation, where each one appears: tests that never execute and reference-side imports that block valid candidates (\S\ref{sec:quanbench_audit}), scoring contracts that check undisclosed quantities (\S\ref{sec:qbe_results}), a judge-scored suite with a non-i.i.d.\ protocol (\S\ref{sec:qcoder_results}), and top-model differences that do not survive multiplicity correction (\S\ref{sec:separability}).

%% file: sections/3_methods.tex
\begin{table}[t]
\centering
\footnotesize
\setlength{\tabcolsep}{5pt}
\begin{tabular}{lcccc}
\toprule
\textbf{Dataset} & \textbf{\# Tasks} & \textbf{Difficulty} & \textbf{Eval Type} & \textbf{Tests?} \\
\midrule
Qiskit (Standard) & 151 & Standard & Execution & \cmark \\
Qiskit (Hard)     & 151 & Hard     & Execution & \cmark \\
QuanBench-44      & 44  & Medium   & Execution & \cmark \\
QuanBench-117     & 117 & Med--Hard& Execution & \cmark \\
QCoder            & 67 & Mixed & LLM Judge & \xmark \\
QuantumBenchEval (QBE)       & 96  & Mixed    & Execution + LLM Judge & \cmark \\
\bottomrule
\end{tabular}
\caption{Summary of quantum code generation benchmarks. Execution-based datasets use unit tests, while QCoder relies on LLM-based evaluation. QBenchEval is execution-graded except for its VQE topic (T3, 15 tasks), which is scored by an LLM rubric judge.}
\label{tab:quantum_datasets}
\end{table}

\section{Methods}
\label{sec:methods}

\subsection{Benchmarking Framework Overview}
Our harness samples tasks from each suite, generates $n{=}5$ completions per task with a shared prompt template and no reranking, voting, or post-processing, and scores them with each suite's native evaluator (unit tests, judge, or task-specific scorer), logging every program and outcome. We use temperature $0.8$ where the provider exposes it. Decoding is not uniform across providers: most models use temperature 0.8, Claude Fable~5.1 and Muse-Glimmer-30B use 1.0, and QCoder output budgets range from 2{,}048 to 32{,}768 tokens (Appendix~\ref{app:provenance}). We therefore report output validity alongside pass@$k$ and treat cross-model gaps as measured under these configurations.

\subsection{Datasets}
We evaluate models across five quantum programming benchmarks spanning two evaluation paradigms: execution-based correctness and LLM-based judgment.

\textbf{QiskitHumanEval (Standard, 151 tasks)} is a quantum adaptation of the HumanEval benchmark, consisting of short programming problems covering core quantum computing primitives such as state preparation, circuit construction, and measurement. Correctness is determined via unit tests.

\textbf{QiskitHumanEval (Hard, 151 tasks)} is a difficulty-enhanced variant of the standard benchmark, requiring deeper compositional reasoning and multi-step circuit design.

\textbf{QuanBench-117 (117 tasks)} is a large-scale benchmark designed to cover a broad spectrum of quantum programming concepts, emphasizing generalization across diverse tasks.

\textbf{QuanBench-44 (44 tasks)} is a smaller subset of QuanBench that enables controlled comparison under reduced task diversity.

\textbf{QCoder (67 tasks)} consists of open-ended quantum programming problems without canonical solutions or executable test cases. These tasks emphasize end-to-end reasoning and algorithmic design rather than test-driven correctness.

\textbf{QuantumBenchEval (QBE, 96 tasks)} is a new benchmark we introduce, spanning six quantum-computing topics with 15--17 tasks each (Table~\ref{tab:qbe-topics}). Five topics are scored by executing the candidate against task-specific assertions; T3 combines execution with a rubric (below).

Unlike classical code benchmarks, correctness in our setting requires alignment with quantum state behavior, including measurement distributions and circuit semantics, making evaluation fundamentally more challenging than traditional unit-test-based code generation. Table~\ref{tab:quantum_datasets} summarizes the datasets used in our evaluation.

Examples of a task from each dataset can be found in Appendix ~\ref{app:examples}.

\subsection{Models}

We evaluate 15 models in the primary benchmark collection, grouped by specialization and deployment setting.

\paragraph{Hosted general-purpose models.}
These comprise GPT-5 via Azure OpenAI \citep{singh2025openai}, GPT-6 Astra
\citep{openai2026astra}, Claude Opus 4.6 \citep{anthropic2026opus46}, Claude
Fable 5.1 \citep{anthropic2026fable51}, Gemini 3 Flash
\citep{google2025gemini3flash}, Gemini 3.6 Flash
\citep{google2026gemini36flash}, and Gemini 3.1 Pro Preview
\citep{google2026gemini31pro}.

\paragraph{General-purpose open-weight models.}
We include LLaMA-3.1-8B \citep{grattafiori2024llama},
Mistral-7B-Instruct-v0.3 \citep{jiang2023mistral7b},
Gemma-3-4B-IT \citep{gemmateam2025gemma3technicalreport}, and
Muse-Glimmer-30B \citep{lu2026museunifiedagenticharness}.
LLaMA-3.1-8B is the base, non-instruction-tuned checkpoint.

\paragraph{Quantum-specialized models.}
We evaluate Qwen2.5-Coder-14B-Qiskit \citep{qiskit2025qwen},
Mistral Small 3.2 24B Qiskit \citep{qiskit2025mistral},
and Granite 3.2 8B Qiskit \citep{qiskit2025granite,dupuis2024qiskit}.
\paragraph{Retrieval-augmented system.}
We additionally evaluate Quantum RAG \citep{11544968}, which combines retrieval with a quantum-code-generation model.


\subsection{Evaluation Protocol}
\paragraph{Pass@$k$ and uncertainty.} For a task with $n{=}5$ samples and $c$ successes we use the unbiased estimator $\widehat{\mathrm{pass}@k}=1-\binom{n-c}{k}/\binom{n}{k}$, averaged over tasks, reporting $k\in\{1,3,5\}$ for the primary suites and $k\in\{1,5\}$ for assertion-scored QBE topics. We give pointwise 95\% intervals from 10{,}000 task-bootstrap replicates, paired task-level model comparisons with Holm correction within each suite (105 pairs; 78 for QuanBench-44 and QCoder; Appendix~\ref{app:stats}), and Spearman and Kendall $\tau_b$ rank agreement across suites. Intervals condition on the recorded samples and evaluators and do not account for unresolved scoring defects; affected outcomes are marked as requiring revalidation.

\paragraph{Judge-scored evaluation.} QCoder provides no executable tests, so each candidate receives a binary PASS/FAIL verdict from an LLM judge (per-run judge identities in Appendix~\ref{app:provenance}). Claude Opus 4.6 is excluded from the QCoder roster to avoid self-evaluation; Claude Fable 5.1 shares a provider with the QCoder and T3 judges and was
retained, so its judge-scored results are flagged given documented
self-preference effects~\citep{zheng2023judging}.

\paragraph{QBE T3: open-ended energy estimation.} T3 contains 15 tasks requiring Qiskit implementations of variational ground-state energy estimation for molecular and lattice systems. Because multiple implementations can be valid, T3 has no binary outcome. Each judged sample is scored on a fixed 10-point rubric: energy accuracy (3), chemical accuracy (1), required outputs (2), qubit and circuit specification (2), and scientific reasoning (2). Candidate code runs in an isolated subprocess; energy accuracy is recomputed deterministically from the executed energy against a task-specific reference, and the remaining criteria are scored by Claude Sonnet 4.6 from the candidate code and execution result. 

\paragraph{Diagnostics.} We report valid-output rates (nonempty and syntactically valid, with total-sample denominators), rule-based classifications of saved evaluation errors, and observed task coverage (empirical pass@5) with pairwise complementarity. Union coverage is an oracle quantity that uses additional generation budget, not a deployable selector. We also compute CodeBLEU \citep{ren2020codebleumethodautomaticevaluation} and its per-sample point-biserial correlation with pass/fail. None of these diagnostics establishes functional or quantum-semantic correctness. Error-label definitions, CodeBLEU scoring details, T1 resource closeness, token usage, and incomplete-run bounds are given in Appendix~\ref{app:diagnostics}.

%% file: sections/4_experiments_ICLR.tex
\section{Experiments}
\label{sec:experiments}

We evaluate current language models across the five primary quantum
code-generation suites described in Section~\ref{sec:methods}, and separately
analyze QuantumBenchEval (QBE) as an evaluator-focused benchmark. Unless
otherwise stated, results are reported as pass@1 with pointwise 95\% task-bootstrap
confidence intervals from 10,000 replicates. We emphasize task-level uncertainty
and paired comparisons rather than interpreting raw leaderboard ordering as
evidence of model separation. QuanBench results are retained as recorded
evaluator outcomes but require replay following the evaluator audit described in
Section~\ref{sec:quanbench_audit}.

\subsection{Overall Performance}
\label{sec:results}

Table~\ref{tab:main_p1} summarizes pass@1 across the five primary suites.
Performance varies substantially across benchmarks, and no single ordering
describes all five evaluations. QiskitHumanEval Standard is particularly
concentrated at the top: GPT-6 Astra reaches 98.7\% pass@1
[96.7--100.0], Gemini 3.1 Pro Preview reaches 97.6\%
[95.1--99.6], and Muse-Glimmer-30B reaches 95.1\%
[92.5--97.5]. The corresponding Hard split remains considerably less
saturated, with the highest recorded pass@1 at 63.8\% [56.4--71.0].

QuanBench exhibits a different ordering. Claude Fable 5.1 records the highest
pass@1 on QuanBench-44 at 60.0\% [45.9--73.6], while several systems cluster
near 84--85\% on QuanBench-117. These QuanBench values should not be interpreted
as validated capability estimates, however, because our evaluator audit identifies
defects affecting both apparent passes and apparent failures
(Section~\ref{sec:quanbench_audit}).

QCoder produces another distinct performance profile. Among complete runs,
Claude Fable 5.1 records 43.0\% pass@1 [33.7--52.2], followed by GPT-6 Astra
at 32.8\% [25.4--40.6]. Because QCoder conditions later generations on prior
attempts, these pass@$k$ values describe the recorded sequential protocol rather
than an unbiased independent-sampling estimator.

\begin{table}[t]
\centering
\scriptsize
\setlength{\tabcolsep}{3.2pt}
\resizebox{\textwidth}{!}{%
\begin{tabular}{lccccc}
\toprule
\textbf{Model} &
\textbf{Standard} &
\textbf{Hard} &
\textbf{QB-44$^\ddagger$} &
\textbf{QB-117$^\ddagger$} &
\textbf{QCoder$^\dagger$} \\
\midrule

\multicolumn{6}{l}{\textit{Quantum-specialized models}} \\
Granite-Qiskit
& 20.5 [15.9--25.6]
& 11.4 [7.7--15.5]
& 10.5 [5.9--15.9]
& 31.3 [26.3--36.2]
& 7.5 [1.5--14.9] \\

Qwen-Qiskit
& 53.8 [47.5--59.9]
& 22.3 [17.0--27.9]
& 27.7 [17.3--39.5]
& 74.5 [67.5--81.0]
& 4.8 [1.2--9.3] \\

Mistral-Qiskit
& 67.2 [60.1--74.2]
& 23.4 [17.7--29.5]
& 25.9 [14.5--37.7]
& 79.8 [73.2--86.3]
& 4.5 [1.2--9.0] \\

\midrule
\multicolumn{6}{l}{\textit{Retrieval-augmented system}} \\
Quantum RAG
& 41.3 [35.6--47.2]
& 8.5 [5.3--11.9]
& 15.9 [7.7--25.0]
& 42.1 [35.2--49.1]
& 9.3 [3.9--15.5] \\

\midrule
\multicolumn{6}{l}{\textit{General-purpose open-weight models}} \\
LLaMA-3.1-8B
& 5.7 [3.7--7.8]
& 0.1 [0.0--0.4]
& 31.4 [22.7--40.0]
& 28.0 [23.2--32.8]
& 4.2 [1.2--8.1] \\

Mistral-7B
& 9.4 [6.1--13.1]
& 0.4 [0.0--0.9]
& 5.0 [0.5--10.9]
& 11.6 [6.7--16.9]
& 2.4 [0.3--5.4] \\

Gemma-3-4B
& 65.2 [57.6--72.6]
& 0.1 [0.0--0.4]
& 11.8 [3.6--21.4]
& 77.6 [70.8--84.1]
& 3.9 [0.0--8.7] \\

Muse-Glimmer-30B
& 95.1 [92.5--97.5]
& 28.3 [22.1--34.7]
& 45.9 [33.2--58.6]
& 81.9 [75.0--88.0]
& --$^{*}$ \\

\midrule
\multicolumn{6}{l}{\textit{Hosted models}} \\
GPT-5
& 46.5 [39.2--53.9]
& 24.1 [18.0--30.5]
& 58.6 [45.5--71.4]
& 81.0 [74.4--87.2]
& 7.8 [2.1--14.6] \\

Gemini 3 Flash
& 62.8 [55.6--69.8]
& 32.6 [26.0--39.5]
& 25.5 [15.9--36.4]
& 44.4 [37.4--51.8]
& 2.4 [0.9--4.2] \\

Gemini 3.6 Flash
& 64.0 [56.6--71.0]
& 40.8 [33.5--48.2]
& 31.8 [20.0--44.5]
& 50.6 [43.1--58.0]
& 23.3 [16.4--30.7] \\

Claude Opus 4.6
& 71.7 [64.4--78.5]
& 44.1 [36.6--51.9]
& 52.3 [38.2--66.4]
& 84.4 [77.9--90.6]
& -- \\

Claude Fable 5.1
& 70.7 [63.4--78.0]
& 56.0 [48.6--63.4]
& 60.0 [45.9--73.6]
& 84.6 [77.8--90.6]
& 43.0 [33.7--52.2] \\

Gemini 3.1 Pro Preview
& 97.6 [95.1--99.6]
& 58.1 [51.1--65.3]
& 57.3 [42.3--70.9]
& 83.1 [76.4--89.6]
& 22.1 [14.6--30.1] \\

GPT-6 Astra
& 98.7 [96.7--100.0]
& 63.8 [56.4--71.0]
& 54.5 [38.6--68.2]
& 84.6 [77.8--90.6]
& 32.8 [25.4--40.6] \\

\bottomrule
\end{tabular}}
\caption{Pass@1 (\%) with pointwise 95\% task-bootstrap confidence intervals
across the five primary suites. $^\dagger$QCoder is judge-scored and uses a
sequential, non-i.i.d.\ generation protocol. $^\ddagger$QuanBench values are
recorded evaluator outcomes requiring revalidation under the corrected isolated
evaluator. $^{*}$Muse QCoder is incomplete (45/67 tasks) and is therefore
excluded from full-dataset comparison. Missing results are shown as ``--''
rather than imputed as zero.}
\label{tab:main_p1}
\end{table}

\subsection{Benchmark Saturation and Statistical Separability}
\label{sec:separability}

The Standard and Hard QiskitHumanEval suites provide a useful contrast in
remaining benchmark headroom. Across five attempts per model, only one of the
151 Standard tasks is unsolved by every evaluated model, compared with 31 of
151 Hard tasks. The highest observed pass@1 correspondingly falls from 98.7\%
on Standard to 63.8\% on Hard. All 15 models record lower pass@1 on the Hard
split, although we treat this as a benchmark-level contrast rather than a causal
estimate of difficulty because task-ID correspondence alone does not establish
a controlled paired intervention.

Raw leaderboard ordering also overstates how strongly the evaluated suites
separate their highest-scoring systems. Table~\ref{tab:separability} reports
the two highest observed pass@1 values on each suite together with paired
task-level differences. After Holm correction within each suite, none of the
five top-two gaps is significant. On Hard and QCoder the unadjusted evidence is
suggestive ($p{=}0.016$ and $0.003$; the paired intervals exclude zero), but it
does not survive correction for the full set of within-suite comparisons. An
observed first place therefore should not be read as evidence that the leading
model is reliably better than the runner-up.

Suite choice also changes the ordering itself. Pass@1 rank agreement is high
among most execution-scored suites (Spearman $\rho$ 0.78--0.89) but only
$\rho{=}0.54$ between QiskitHumanEval-Standard and QCoder
(Appendix~\ref{app:stats}), so no single suite recovers the cross-suite
ranking. Additional samples do not separate the leaders either: the strongest
models gain at most about 6\,pp from pass@1 to pass@5 on the execution-scored
suites (zero for GPT-6 Astra on three of four), whereas quantum-specialized
models gain up to 38\,pp and reorder the lower half of the leaderboard
(Appendix~\ref{app:coverage}).

\begin{table}[t]
\centering
\scriptsize
\setlength{\tabcolsep}{3.5pt}
\resizebox{\textwidth}{!}{%
\begin{tabular}{lllllll}
\toprule
\textbf{Suite} &
\textbf{Highest observed} &
\textbf{P@1 [95\% CI]} &
\textbf{Runner-up} &
\textbf{Paired gap, pp [95\% CI]} &
\textbf{$p$} &
\textbf{Holm $p$} \\
\midrule
Standard & GPT-6 Astra & 98.7 [96.7--100.0] & Gemini 3.1 Pro & 1.1 [0.1--2.6] & 0.128 & 0.90 \\
Hard & GPT-6 Astra & 63.8 [56.4--71.0] & Gemini 3.1 Pro & 5.7 [1.3--10.3] & 0.016 & 0.23 \\
QuanBench-44$^\ddagger$ & Claude Fable 5.1 & 60.0 [45.9--73.6] & GPT-5$^{\diamond}$ & 1.4 [$-$8.6--10.9] & 0.860 & 1.00 \\
QuanBench-117$^\ddagger$ & Claude Fable 5.1 & 84.6 [77.8--90.6] & GPT-6 Astra (tie) & 0.0 [0.0--0.0] & 1.000 & 1.00 \\
QCoder$^\dagger$ & Claude Fable 5.1$^{\S}$ & 43.0 [33.7--52.2] & GPT-6 Astra & 10.1 [3.9--16.7] & 0.003 & 0.13 \\
\bottomrule
\end{tabular}}
\caption{Top-two pass@1 comparison per suite. The paired gap is the mean
per-task difference between the two models, with a pointwise task-bootstrap
interval; $p$ is a two-sided paired permutation test (10{,}000 sign flips).
Holm adjustment is applied within each suite over all model pairs (105; 78 for QuanBench-44 and QCoder).
Neither the intervals nor the unadjusted $p$ values are
multiplicity-adjusted. $^\dagger$Judge-scored, sequential protocol.
$^\ddagger$Recorded evaluator outcomes pending replay. $^{\S}$Judge shares a
provider with the evaluated model. $^{\diamond}$High empty-output rate
(Appendix~\ref{app:validity}).}
\label{tab:separability}
\end{table}

Observed coverage reinforces the difference between the suites without implying
that any task is intrinsically unsolvable. Across the available roster, the
number of tasks unsolved in all five recorded attempts is 1/151 for Standard,
31/151 for Hard, 9/44 for QuanBench-44, 17/117 for QuanBench-117, and 12/67 for
QCoder. These are empirical coverage statistics, equivalent to observed
pass@5 coverage across the roster, rather than estimates of a fundamental
capability ceiling.

\subsection{Category-Level Performance}
\label{sec:categories}

\begin{figure}[t]
    \centering
    \includegraphics[width=.9\textwidth]{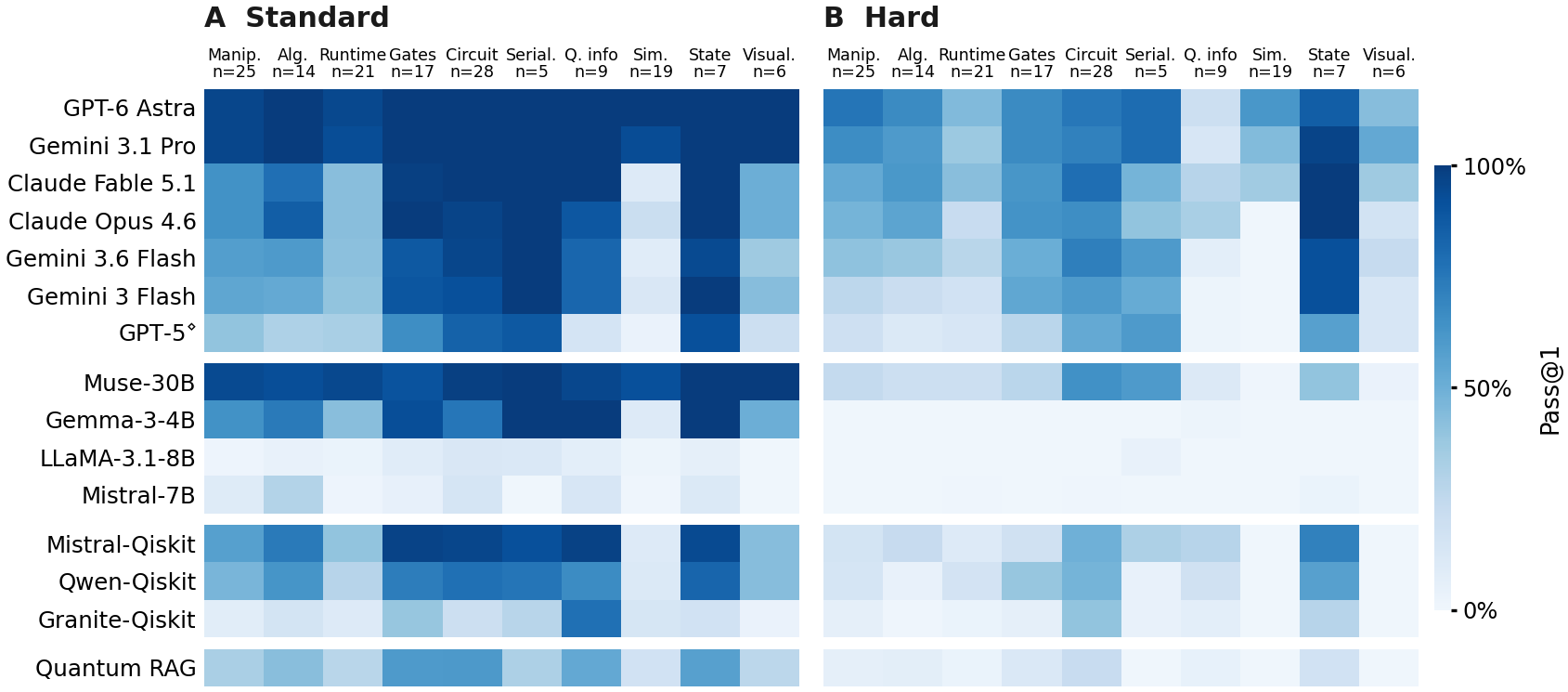}
    \caption{\textbf{Category-level pass@1 on QiskitHumanEval Standard and Hard.}
    Rows are models, grouped by family; columns are the ten task categories, with
    per-category task counts ($n$) in the headers. Standard is near saturation for
    the strongest models across most categories, while Hard shows larger
    differences across task types. Categories with fewer than ten tasks (e.g.,
    Serialization, $n{=}5$) have wide intervals, and a single-category difference
    should not be read as a model ranking. Per-category 95\% task-bootstrap
    intervals for every model are given in Appendix~\ref{app:categories}.}
    \label{fig:category_breakdown}
\end{figure}

Figure~\ref{fig:category_breakdown} reports pass@1 across the ten shared
QiskitHumanEval task categories for Standard and Hard. Standard is broadly
saturated among the strongest models: GPT-6 Astra reaches 100\% pass@1 in
eight of ten categories, with the remaining two at 96.0\% and 95.2\%.

Hard reveals substantially greater category-level variation. For Astra,
pass@1 ranges from 20.0\% on \textsc{Quantum Information} and 44.8\% on
\textsc{Backend and Runtime} to 75.0\% on \textsc{Quantum Circuit Generation}
and 85.7\% on \textsc{State Preparation and Analysis}; similar heterogeneity
appears across other strong models. Because category sizes range from 5 to
28 tasks, these results are interpreted descriptively rather than as a
secondary model ranking.

\begin{figure}[t]
    \centering
    \includegraphics[width=\textwidth]{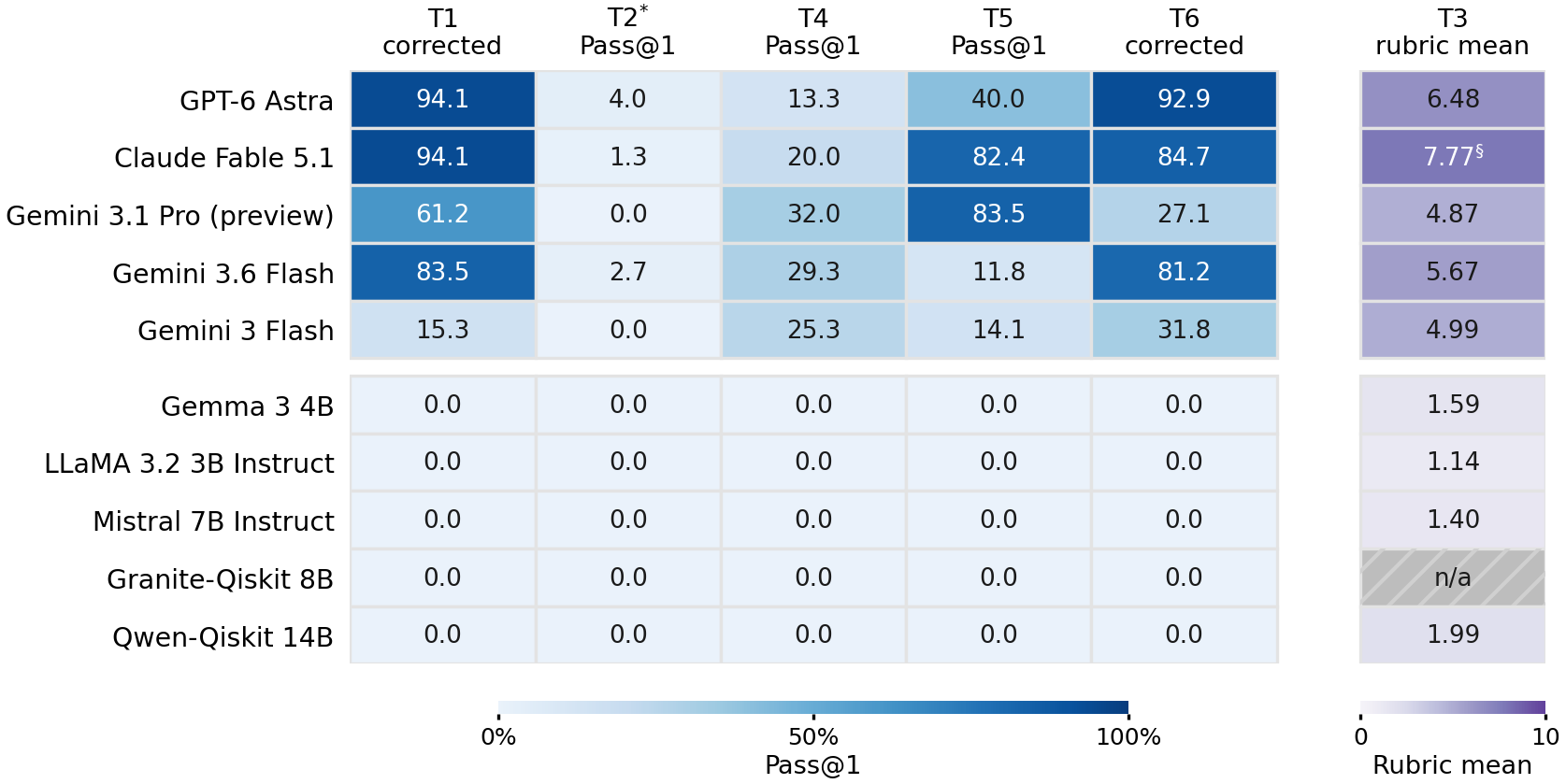}
    \caption{\textbf{QuantumBenchEval results by topic.} Pass@1 (\%) for T1, T2, T4, T5 and T6 (blue, shared scale). T3 (purple, separate scale) is an LLM-judged rubric mean over judged samples only; it is not a pass rate and should not be compared with the other columns. n/a: no judged samples. T1 and T6 use corrected scoring that removes checks on quantities the prompts did not specify, applied to the same saved outputs. This raises GPT-6 Astra from 0\% to 94.1\% on T1 and from 1.2\% to 92.9\% on T6. $^{*}$T2 is reported as recorded (Appendix~\ref{app:qbe-rescore}). $^{\S}$The judge shares a provider with the evaluated model. The horizontal gap separates hosted frontier models from small open-weight and quantum-specialized models.}

    \label{fig:qbe_breakdown}
\end{figure}

\subsection{QuantumBenchEval: Challenging Tasks and Evaluation Sensitivity}
\label{sec:qbe_results}

QuantumBenchEval (QBE) complements the established suites with 96 new tasks
spanning six quantum-computing topics. Instead of short API-completion problems,
its tasks are multi-step scientific workflows that require reasoning about
algorithm design, resource use, optimization, simulation, and scientific
outputs. Each task is scored by its own evaluator, and because the tasks are
newly constructed, they are less likely than long-standing public suites to
appear in model training data.

Figure~\ref{fig:qbe_breakdown} shows substantial variation across both models
and topics. The hosted frontier models score above zero on several topics, while
the smaller open-weight and quantum-specialized systems score zero pass@1 on
every assertion-scored topic. We do not read these zeros as an absence of
quantum knowledge. The same models receive nonzero T3 rubric scores (1.1--2.0
of 10), so the zeros more plausibly reflect failure to produce complete outputs
in the required structure than failure to engage with the task. Among the
stronger systems, performance is highly topic dependent. GPT-6 Astra and Claude
Fable~5.1 both reach 94.1\% pass@1 on corrected T1, yet score 40.0\%
[22.4--58.8] and 82.4\% [69.4--94.1] on T5. Gemini 3.1 Pro Preview reaches
83.5\% [76.5--90.6] on T5, while no model exceeds 32\% on T4. The six topics
therefore probe different capabilities, and no single model ordering holds
across them.

We applied the same audit to our own evaluators, checking whether each scoring rule matches the information given to the model. It flagged three topics. T1's evaluator checked exact resource quantities the prompts never specified, and T6's required a scalar \texttt{reproducibility} value whose form the prompts never stated. Removing only these checks, while keeping every correctness check and rescoring the saved outputs, raises GPT-6 Astra from 0\% to 94.1\% on T1 and from 1.2\% to 92.9\% on T6; on T6 this moves it from last to first among the frontier models. All 32 reference solutions pass under both evaluators, and no replayed sample changes from pass to fail (Appendix~\ref{app:qbe-rescore}). T2 also checks undisclosed evaluation counts, but removing them changes little (at most 13.3\% pass@1 on the 12 internally consistent tasks), and three T2 tasks contain data that contradict their required answers, so we report T2 as recorded. T3 is scored differently again, as a rubric mean over judged samples rather than pass@1.

\subsection{QuanBench Evaluator Audit}
\label{sec:quanbench_audit}

The 17 QuanBench-117 tasks unsolved across the observed roster initially appear
to form a shared performance ceiling. Inspection instead identifies a common
benchmark-side obstruction: for each of the 17 tasks, the dominant recorded
failure corresponds to an import in benchmark-owned reference or test code,
including dependencies from \texttt{qiskit\_algorithms},
\texttt{qiskit\_optimization}, \texttt{qiskit.primitives}, and
\texttt{qiskit\_machine\_learning}.This does not imply that every candidate on these tasks is correct (some outputs fail before reaching the shared obstruction), but it prevents some outputs fail before reaching the shared
obstruction, but it prevents the observed ceiling from being interpreted as a
pure model-capability limit.

More importantly, source inspection and controlled synthetic probes identify
three defects in the current QuanBench evaluator. First, the evaluator executes
test scripts without explicitly loading and running their \texttt{unittest}
test cases; all 117 current QuanBench-117 scripts contain test classes and no
top-level expression calls. A deliberately incorrect candidate therefore passes
when the failing assertion is confined to such a test class. Second, canonical
code is executed in the candidate namespace before generated code, allowing an
empty candidate to satisfy an explicit assertion by inheriting the canonical
function. Third, a valid synthetic candidate can fail before execution when the
canonical solution imports a missing dependency. The false-pass mechanism is not 
only hypothetical: a recorded QuanBench-117 pass (Task~110, Claude Opus~4.6) 
returns a circuit that is not equivalent to CNOT up to global phase (Appendix~\ref{app:qualitative}).

These controlled probes demonstrate mechanisms for both false passes and false
failures. Historical evaluator source was not immutably captured, so they do not
prove that every saved QuanBench outcome was produced by the exact inspected
revision. We therefore retain the historical scores as \emph{recorded evaluator
outcomes} and require replay of unchanged candidates under a versioned,
isolated evaluator before interpreting either successes or failures as validated
capability estimates. QuanBench-44 has a different test-script structure, so the
class-only finding for QuanBench-117 is not automatically extended to every
QuanBench-44 task.

\subsection{Judge-Scored QCoder}
\label{sec:qcoder_results}

QCoder evaluates open-ended programs with a model-based judge rather than
executable unit tests. Among complete evaluations, Claude Fable~5.1$^{\S}$
records 43.0\% pass@1 [33.7--52.2] and 67.2\% observed pass@5 [55.2--77.6],
while GPT-6 Astra records 32.8\% pass@1 [25.4--40.6] and the same 67.2\%
observed pass@5. Equal coverage does not mean identical behavior: each model
solves 45/67 tasks within five attempts, with 40 shared successes and five
unique successes each, so their oracle union covers 50/67 tasks (74.6\%). This
shows complementary observed coverage, not a demonstrated ensemble, since
combining the systems doubles the generation budget and relies on oracle
knowledge of which candidate succeeds.

These results are also conditioned on the generation configuration. GPT-6 Astra,
Claude Fable~5.1, and Muse-Glimmer-30B produced no usable code under a
2{,}048-token output limit, so their runs used 32{,}768 tokens. Earlier runs,
including GPT-5, used 2{,}048, and GPT-5's 63\% empty-output rate on QCoder
shows that the cap was binding for it as well (Appendix~\ref{app:validity}).
The judge also changed between the two groups (\texttt{claude-sonnet-4} before
April 2026, \texttt{claude-sonnet-4-6} after), and neither group was re-judged
with the other judge. Gaps between the earlier and later runs therefore reflect
model capability together with output budget and judge version, and the
within-group comparison of Fable and Astra is the most controlled one QCoder
provides.

Muse-Glimmer-30B is provisional: it has completed 45/67 tasks (completed-subset
pass@1 53.3\%; full-dataset bounds 35.8--68.7\%, obtained by assigning every
unfinished task failure or success) and is excluded from QCoder rankings.

%% file: sections/5_discussions.tex
\section{Discussion and Conclusion}
Our results show that measured quantum code-generation performance depends as much on how correctness is operationalized as on the model evaluated. Correcting one QBE evaluator changes two models' pass@1 by 94.1\,pp on unchanged outputs; the current QuanBench evaluator admits both false passes and false failures; and no primary suite separates its top two models after multiplicity correction. Meanwhile QiskitHumanEval-Standard is near saturation, so headroom now lies in harder or less-exposed tasks such as the Hard split and QBE.
\paragraph{Implications for model comparison.} These effects compound. A leaderboard entry for QuanBench-117 reflects the model, its output budget, whether the evaluator runs its tests, and whether reference code leaks into the candidate namespace; a QCoder entry additionally reflects the judge version. Comparisons are most trustworthy within a single configuration and evaluator revision, as with Claude Fable~5.1 and GPT-6 Astra on QCoder. Across configurations, differences smaller than the evaluator effects documented here should not be interpreted as capability differences.
\paragraph{Recommendations.} Quantum code benchmarks should (i) validate reference solutions and confirm that tests execute under the released harness; (ii) isolate candidate code from reference code; (iii) disclose every quantity the scorer checks; (iv) report valid-output rates alongside pass@$k$; (v) report uncertainty and corrected paired comparisons rather than raw ordering; and (vi) use independent sampling whenever pass@$k$ is estimated.
\paragraph{Limitations.} QuanBench outcomes await replay under an isolated evaluator. QBE T2 contracts are corrected only in the release, and T1/T6 rescoring used replay on a different platform (5 of 1{,}600 samples differ). QCoder uses a non-i.i.d.\ protocol, its judge changed between runs, and all judges share a provider with one evaluated model. Decoding settings and output budgets differ across models, and our roster reflects one point in time.

%% file: sections/6_appendix_ICLR.tex
\appendix
\sloppy 

\section{Evaluation Details}
\label{app:eval}

\subsection{Prompting Setup}
\label{app:prompt}

Each task is presented as a natural-language instruction together with the
suite's function signature or code template, and the model is asked for a
complete implementation. QiskitHumanEval and QuanBench use the template below.
QCoder and QuantumBenchEval supply their own problem statements and required
function formats verbatim (examples in Appendices~\ref{app:examples}
and~\ref{app:qbe-examples}).

\begin{lstlisting}[style=quantumcode]
You are an expert in quantum computing and Qiskit.

Given the following problem, generate a correct and executable
Python function using Qiskit.

Problem:
{problem_description}

Requirements:
- Return only valid Python code
- Do not include explanations
- Follow the specified function signature exactly
- Ensure the code passes all tests

Answer:
\end{lstlisting}

We draw $n{=}5$ samples per task. Decoding settings are not identical across
providers; the settings actually applied are recorded in
Appendix~\ref{app:provenance}.

\subsection{Run Provenance}
\label{app:provenance}

Table~\ref{tab:provenance} records the identifier, access route, decoding
settings, and generation dates for each model, and Table~\ref{tab:judges}
records the judge used for each QCoder run. Three differences bear on
cross-model comparison. First, the maximum output budget for QCoder differs by
a factor of 16 between earlier runs (2{,}048 tokens) and the most recent
reasoning-model runs (32{,}768 tokens). Second, Claude Fable~5.1 and
Muse-Glimmer-30B were sampled at temperature 1.0 rather than 0.8. Third, the
QCoder judge changed from \texttt{claude-sonnet-4-20250514} (runs before April
2026) to \texttt{claude-sonnet-4-6} (runs from September 2026), and no run was
re-judged with the other judge. QCoder scores from the two periods are
therefore not strictly comparable.

\begin{table}[ht]
\centering
\scriptsize
\setlength{\tabcolsep}{3pt}
\resizebox{\linewidth}{!}{%
\begin{tabular}{lllccl}
\toprule
Model & Identifier & Access & Temperature & Max output tokens & Generated \\
\midrule
GPT-5 & \texttt{gpt-5} & Azure OpenAI & 0.8 & 2048$^\dagger$ & 2025-12-16 -- 2026-03-29 \\
GPT-6 Astra & \texttt{gpt-6-astra} & OpenAI API & 0.8 & 32768$^\dagger$ & 2026-09-22 -- 2026-09-23 \\
Claude Opus 4.6 & \texttt{claude-opus-4-6} & Anthropic API & 0.8 & 2048 & 2026-02-09 \\
Claude Fable 5.1 & \texttt{claude-fable-5-1} & Anthropic API & 1.0 & 32768$^\dagger$ & 2026-09-10 -- 2026-09-16 \\
Gemini 3 Flash & \texttt{gemini-3-flash-preview} & Gemini API & 0.8 & 2048$^\dagger$ & 2026-02-09 -- 2026-03-31 \\
Gemini 3.6 Flash & \texttt{gemini-3.6-flash} & Gemini API & 0.8 & 2048$^\dagger$ & 2026-09-09 \\
Gemini 3.1 Pro Preview & \texttt{gemini-3.1-pro-preview} & Gemini API & 0.8 & 2048$^\dagger$ & 2026-09-10 \\
LLaMA-3.1-8B (base) & \texttt{meta-llama/Llama-3.1-8B} & local & 0.8 & 2048$^\dagger$ & 2026-01-06 -- 2026-09-09 \\
Mistral-7B-Instruct-v0.3 & \texttt{mistralai/Mistral-7B-Instruct-v0.3} & local & 0.8 & 2048$^\dagger$ & 2026-01-12 -- 2026-03-31 \\
Gemma-3-4B-IT & \texttt{google/gemma-3-4b-it} & local & 0.8 & 2048$^\dagger$ & 2026-01-14 -- 2026-03-30 \\
Muse-Glimmer-30B & \texttt{meta-models/Muse-Glimmer-30B} & local & 1.0 &32768$^\dagger$ & 2026-09-21 \\
Granite-3.2-8B-Qiskit & \texttt{qiskit/granite-8b-qiskit} & local & 0.8 & 2048$^\dagger$ & 2025-12-16 -- 2026-03-31 \\
Mistral-Small-3.2-24B-Qiskit & \texttt{mistral-small-3.2-24b-qiskit} & local & 0.8 & 2048$^\dagger$ & 2026-02-09 -- 2026-03-31 \\
Qwen2.5-Coder-14B-Qiskit & \texttt{qwen2.5-coder-14b-qiskit} & local & 0.8 & 2048$^\dagger$ & 2026-02-12 -- 2026-03-30 \\
Quantum RAG & \texttt{quantum-rag}  & local & 0.8 & 2048$^\dagger$ & 2026-02-13 -- 2026-09-10 \\
\bottomrule
\end{tabular}}
\\[2pt]
{\scriptsize $^\dagger$QCoder runs only; the other suites' output-token limits are not recorded in the run metadata. All runs use $n=5$ samples per task.}
\caption{Generation provenance.}
\label{tab:provenance}
\end{table}

\begin{table}[ht]
\centering
\scriptsize
\setlength{\tabcolsep}{4pt}
\begin{tabular}{llll}
\toprule
Model & Suite & Judge & Run timestamp \\
\midrule
GPT-5 & QCoder & \texttt{claude-sonnet-4-20250514} & 2026-03-29 12:05 \\
Gemma-3-4B-IT & QCoder & \texttt{claude-sonnet-4-20250514} & 2026-03-30 18:22 \\
Qwen2.5-Coder-14B-Qiskit & QCoder & \texttt{claude-sonnet-4-20250514} & 2026-03-30 22:22 \\
Granite-3.2-8B-Qiskit & QCoder & \texttt{claude-sonnet-4-20250514} & 2026-03-31 06:44 \\
Gemini 3 Flash & QCoder & \texttt{claude-sonnet-4-20250514} & 2026-03-31 12:13 \\
Mistral-Small-3.2-24B-Qiskit & QCoder & \texttt{claude-sonnet-4-20250514} & 2026-03-31 14:54 \\
Mistral-7B-Instruct-v0.3 & QCoder & \texttt{claude-sonnet-4-20250514} & 2026-03-31 15:41 \\
LLaMA-3.1-8B (base) & QCoder & \texttt{claude-sonnet-4-6} & 2026-09-09 00:49 \\
Gemini 3.6 Flash & QCoder & \texttt{claude-sonnet-4-6} & 2026-09-09 17:41 \\
Quantum RAG & QCoder & \texttt{claude-sonnet-4-6} & 2026-09-10 11:27 \\
Gemini 3.1 Pro Preview & QCoder & \texttt{claude-sonnet-4-6} & 2026-09-10 17:25 \\
Claude Fable 5.1 & QCoder & \texttt{claude-sonnet-4-6} & 2026-09-16 10:14 \\
GPT-6 Astra & QCoder & \texttt{claude-sonnet-4-6} & 2026-09-23 04:39 \\
\bottomrule
\end{tabular}
\caption{Judge used for each judge-scored run. QCoder is the only judge-scored suite.
Muse-Glimmer-30B and Claude Opus 4.6 have no QCoder run.}
\label{tab:judges}
\end{table}

\subsection{Diagnostic Definitions}
\label{app:diagnostics}

\paragraph{CodeBLEU.} For suites with canonical solutions, we compute CodeBLEU
\citep{ren2020codebleumethodautomaticevaluation} with equal weights on its four
components. Main-table values are corpus-level scores over sample index zero for
each task, with empty generations kept as empty predictions. We score only caches
matched to the selected model, benchmark split, and ordered evaluation outcomes;
unmatched runs are marked unavailable. Per-sample point-biserial correlations
with pass/fail use task-cluster bootstrap intervals.

\paragraph{Coverage and complementarity.} Observed task coverage is the fraction
of tasks solved at least once among the five recorded samples, i.e.\ empirical
pass@5. Pairwise intersections and unions of solved-task sets quantify
complementarity; union coverage is an oracle quantity requiring additional
generation budget.

\paragraph{Error labels.} Saved evaluation errors are classified by rule into
syntax errors, possible API-version incompatibilities, missing imports,
keyword/signature mismatches, assertion failures, and timeouts. These labels are
diagnostic. They are not independently validated causes, nor evidence of
training-data staleness.

\paragraph{T1 resource closeness.} For corrected-T1 passing samples we compare
self-reported resource use with reference settings:
\[
C_r=\frac{\min(a_r,t_r)}{\max(a_r,t_r)},\qquad
C_{\mathrm{combined}}=\frac{C_{\mathrm{shots}}+C_{\mathrm{calls}}}{2},
\]
where $a_r$ and $t_r$ are the reported and reference values, averaged over
passing samples, with the count always reported. Closeness does not affect
correctness scoring. The reference targets were not disclosed to candidates, so
it is neither an instruction-following nor a quality metric, and the inputs are
self-reported rather than measured.

\paragraph{Token usage.} Where usage metadata exist, we report input and output
tokens, the coverage of the usage records, and recorded output tokens per
successful sample. These quantities exclude unlogged attempts and do not
establish dollar cost or latency.

\paragraph{Incomplete runs.} Muse-Glimmer-30B's QCoder run covers 45 of 67
tasks. We report completed-subset scores and full-dataset bounds obtained by
assigning every unfinished task failure or success. These bounds are not
confidence intervals or imputations.

\subsection{Statistical Procedures}
\label{app:stats}

\paragraph{Intervals.} Pass@$k$ intervals are percentile intervals from 10{,}000
task-bootstrap replicates: tasks are resampled with replacement, and all samples
of a resampled task are kept.

\paragraph{Paired comparisons.} For each suite and metric, we compare every model
pair on their shared tasks. We report the mean paired difference with a
task-bootstrap interval and a two-sided paired permutation $p$-value from 10{,}000
sign flips, whose resolution floor is $1/10{,}001\approx10^{-4}$. Holm correction
is applied within each suite--metric family (105 pairs for 15 models; 78 for
QuanBench-44, 13 valid models, and for QCoder). We do not pool all comparisons
into one family: at this permutation resolution the smallest attainable pooled
Holm-adjusted $p$ is about $0.15$, so no difference, however large, could reach
significance. Under the per-suite families, 340 of the 471 pass@1 comparisons are
significant at $\alpha{=}0.05$. The top-two gaps in Table~\ref{tab:separability}
are not significant on any suite (Holm-adjusted $p$: Standard 0.90, Hard 0.23,
QuanBench-44 1.00, QuanBench-117 1.00, QCoder 0.13).

\begin{figure}[ht]
\centering
\includegraphics[width=\linewidth]{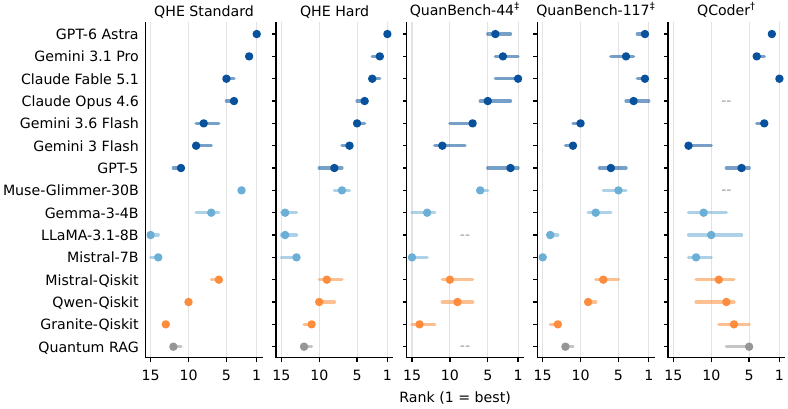}
\caption{Pass@1 rank with 95\% task-bootstrap intervals by suite. Colors: hosted
(dark blue), general open-weight (light blue), quantum-specialized (orange),
retrieval-augmented (gray). ``--'': no valid result.
$^{\ddagger}$QuanBench outcomes await evaluator replay. $^{\dagger}$Judge-scored,
sequential protocol.}
\label{fig:rank-ci}
\end{figure}

\paragraph{Rank intervals.} Figure~\ref{fig:rank-ci} shows each model's observed
pass@1 rank with a 95\% task-bootstrap interval. The intervals are widest on
QuanBench-44, the smallest execution-scored suite, where the rank intervals of
the top five models overlap.

\begin{figure}[ht]
\centering
\includegraphics[width=.62\linewidth]{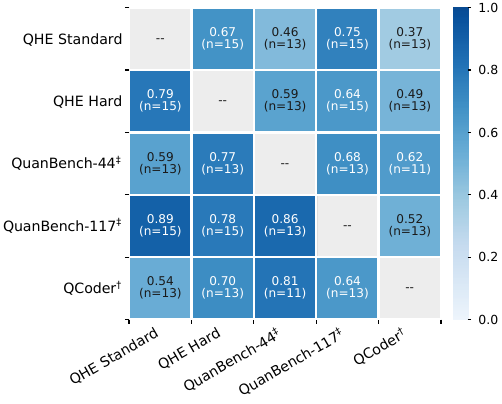}
\caption{Cross-suite rank agreement on pass@1. Lower triangle: Spearman $\rho$;
upper triangle: Kendall $\tau_b$; $n$ = models with valid results on both
suites.}
\label{fig:rank-agreement}
\end{figure}

\paragraph{Cross-suite rank agreement.} Figure~\ref{fig:rank-agreement} reports
Spearman $\rho$ (lower triangle) and Kendall $\tau_b$ (upper triangle) between
per-suite pass@1 across models present in both suites. Agreement is high among the
execution-scored suites (Spearman $\rho$ between 0.78 and 0.89, except
Standard--QuanBench-44 at 0.59) and lowest between QiskitHumanEval-Standard and
QCoder ($\rho{=}0.54$, $\tau_b{=}0.37$). A single-suite leaderboard therefore
recovers only part of the cross-suite ordering. The high Standard--QuanBench-117
agreement ($\rho{=}0.89$) should be read cautiously, because both suites are
compressed near their ceilings.

\section{Primary Suites}
\label{app:primary}

\subsection{Example Problems}
\label{app:examples}

The following examples illustrate the task interfaces and requested behaviors
of the five primary suites. Descriptions are condensed from the repository
prompts; task identifiers and function names are retained. The matched
Standard and Hard examples illustrate a difference in prompt scaffolding,
not a change in the underlying task. QuantumBenchEval is presented separately.

\paragraph{QiskitHumanEval Standard: circuit construction.}
\begin{quote}
\textbf{Task ID:} \texttt{qiskitHumanEval/0}.

\textbf{Task:} Generate a quantum circuit containing a specified number of qubits.

\textbf{Function signature:}
\texttt{def create\_quantum\_circuit(n\_qubits):}

\textbf{Description:}
Return a Qiskit quantum circuit with \texttt{n\_qubits} qubits.
The Standard prompt supplies the \texttt{QuantumCircuit} import and a
function skeleton with a docstring describing the task.

\textbf{Recorded test contract:}
The test calls the function with \texttt{n\_qubits=3} and checks that the
result is a \texttt{QuantumCircuit} containing three qubits.
\end{quote}

\paragraph{QiskitHumanEval Hard: circuit construction without the supplied skeleton.}
\begin{quote}
\textbf{Task ID:} \texttt{qiskitHumanEval/0}.

\textbf{Task:} Generate a quantum circuit containing a specified number of qubits.

\textbf{Required interface:}
\texttt{def create\_quantum\_circuit(n\_qubits):}

\textbf{Description:}
The Hard prompt states the task in natural language and specifies the function
name and argument. It does not supply the import or Python function skeleton
provided by the matched Standard prompt; the model must produce these itself.

\textbf{Recorded test contract:}
For this matched example, the checks remain the same: the returned object must
be a \texttt{QuantumCircuit} with three qubits when called with argument three.
Thus, this example does not establish that Hard uses more demanding semantic
checks.
\end{quote}

\paragraph{QuanBench-44: Grover search.}
\begin{quote}
\textbf{Task ID:} \texttt{01}.

\textbf{Task:} Implement Grover search for the marked state $\lvert 00\rangle$.

\textbf{Function signature:}
\texttt{def grover\_search\_oracle\_00() -> QuantumCircuit:}

\textbf{Description:}
Construct a Qiskit circuit implementing Grover search for
$\lvert 00\rangle$ and return the circuit with measurements.

\textbf{Intended checks in the saved test script:}
The test requests a comparison against the reference circuit using a
KL-divergence helper, with a value in $[0,0.05)$, and checks that
\texttt{00} is the most frequent measurement outcome.
These describe the saved assertions, not independently validated enforcement
by the historical evaluator.
\end{quote}

\paragraph{QuanBench-117: three-qubit state preparation.}
\begin{quote}
\textbf{Task ID:} \texttt{02}.

\textbf{Task:} Prepare the state
\[
    \lvert\psi\rangle = \frac{\lvert 011\rangle-\lvert 100\rangle}{\sqrt{2}}.
\]

\textbf{Function signature:}
\texttt{def generate\_quantum\_state\_qubit3() -> QuantumCircuit:}

\textbf{Description:}
Construct a Qiskit circuit that prepares the specified state and return the
circuit with measurements.

\textbf{Intended checks in the saved test script:}
Measurement outcomes must be restricted to \texttt{011} and \texttt{100},
with each observed frequency within $0.1$ of $0.5$. A separate assertion
requests a phase-helper value of $\pi$ for \texttt{100}, within $0.05$.
The measurement-frequency check alone cannot establish the required relative
phase.
\end{quote}

For both QuanBench examples, the current evaluator audit found that executing
these test-class definitions does not run their test methods. Reference-code
leakage and reference-import failures provide additional failure mechanisms;
the KL-divergence helpers also return a constant zero in the audited source.
Accordingly, the checks above document intended test behavior, and should not
be read as evidence that historical pass labels validate the requested
quantum behavior. Historical outcomes require corrected, isolated replay.

\paragraph{QCoder: single-qubit plus-state preparation.}
\begin{quote}
\textbf{Task ID:} \texttt{QPC001\_A1}.

\textbf{Task:} Prepare $\lvert +\rangle$ from $\lvert 0\rangle$, where
\[
    \lvert +\rangle = \frac{\lvert 0\rangle+\lvert 1\rangle}{\sqrt{2}}.
\]

\textbf{Function signature:}
\texttt{def solve() -> QuantumCircuit:}

\textbf{Description:}
The supplied template imports \texttt{QuantumCircuit}, initializes
\texttt{qc = QuantumCircuit(1)}, and asks the model to insert the state
preparation operations before returning \texttt{qc}.
The problem statement specifies that global phase is ignored by the judge.

\textbf{Evaluation in this study:}
Reported QCoder outcomes are model-judge assessments under the recorded
protocol, rather than independently verified execution-based state comparisons.
\end{quote}

\subsection{Category Composition}
\label{app:categories}

\begin{figure}[ht]
\centering
\includegraphics[width=.75\linewidth]{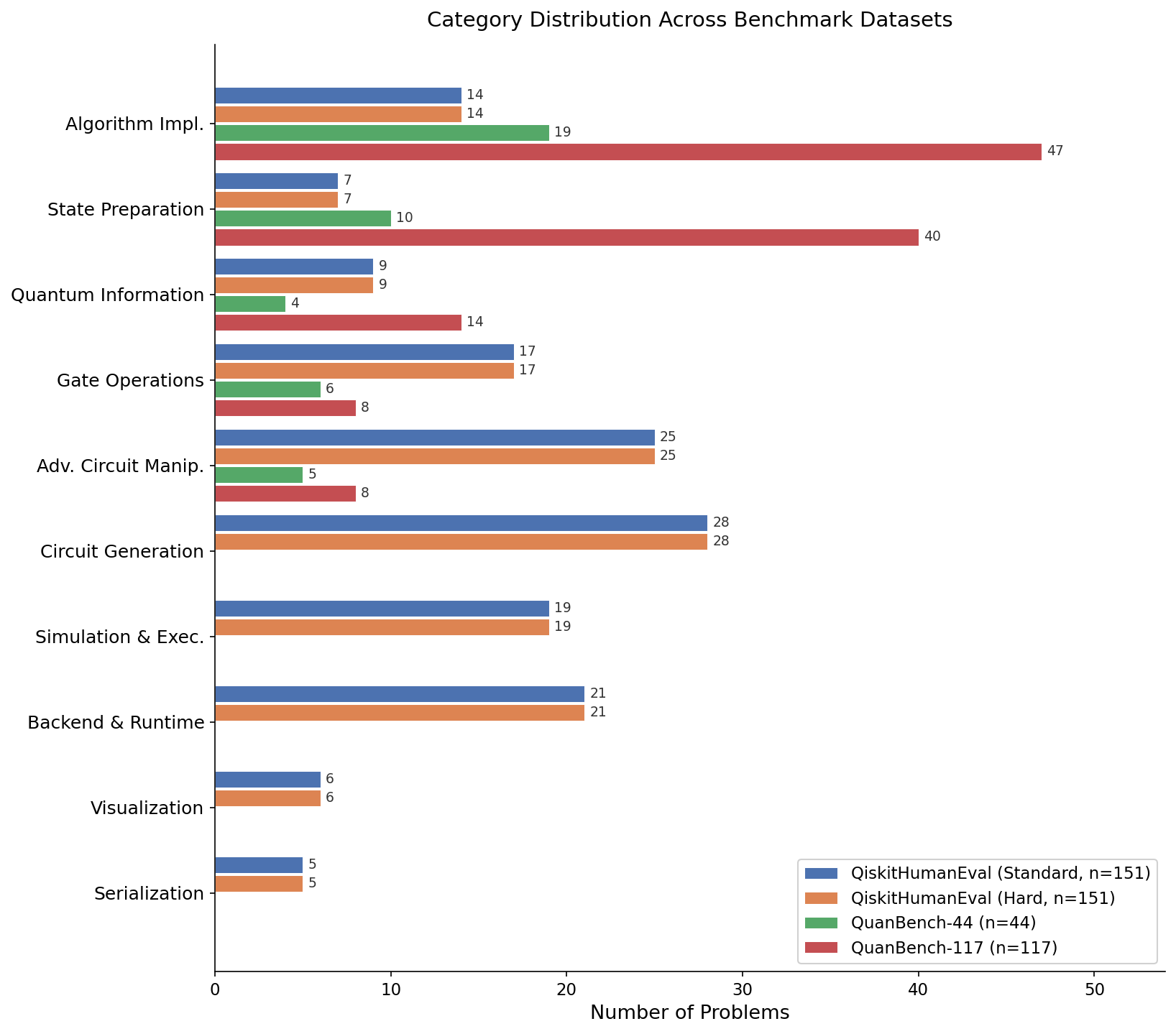}
\caption{Task categories by suite.}
\label{fig:catbreakdown}
\end{figure}

Figure~\ref{fig:catbreakdown} shows how the ten task categories are distributed
across suites. QiskitHumanEval spans a broader portion of the Qiskit ecosystem,
including Backend \& Runtime, Simulation \& Execution, Circuit Generation,
Visualization, and Serialization, none of which appear in QuanBench.
QuanBench-117 is concentrated in Algorithm Implementation (47/117, 40\%) and
State Preparation (40/117, 34\%). Because composition differs this much,
suite-level scores are not comparable measures of a single difficulty scale.

\begin{figure}[p]
\centering
\includegraphics[width=.75\linewidth]{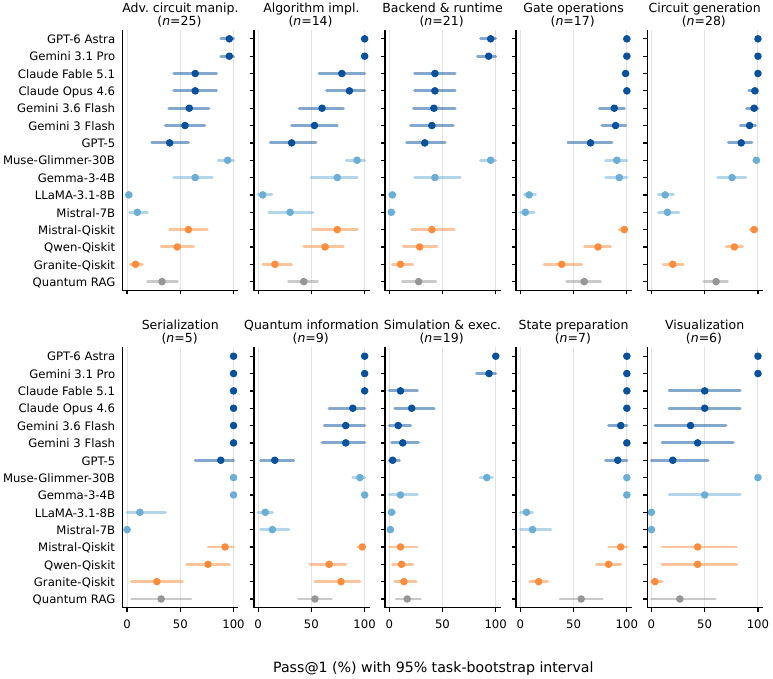}
\caption{Per-category pass@1 on QiskitHumanEval-Standard with 95\% task-bootstrap
intervals ($n$ = tasks in category).}
\label{fig:cat-std}
\end{figure}

\begin{figure}[p]
\centering
\includegraphics[width=.75\linewidth]{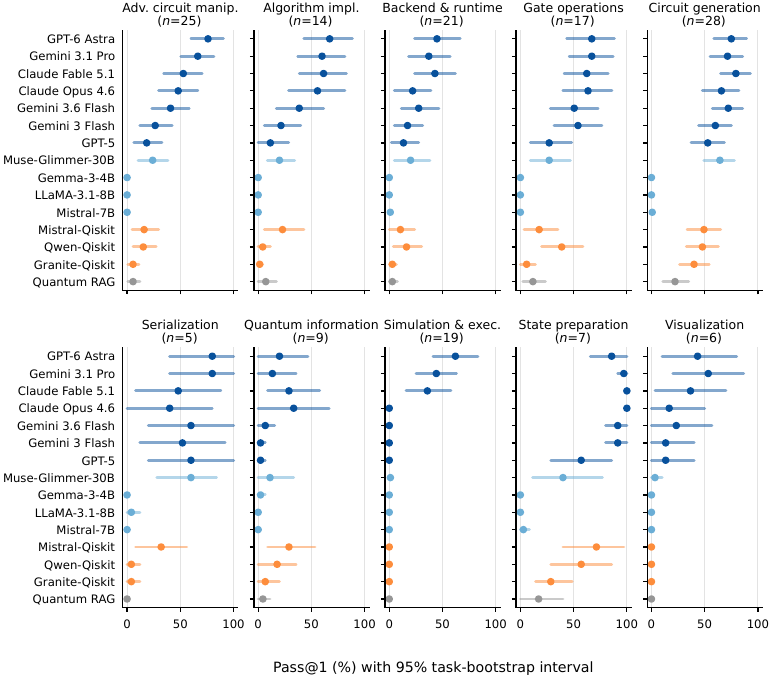}
\caption{Per-category pass@1 on QiskitHumanEval-Hard with 95\% task-bootstrap
intervals.}
\label{fig:cat-hard}
\end{figure}

Figures~\ref{fig:cat-std} and~\ref{fig:cat-hard} give per-category pass@1 with
95\% task-bootstrap intervals for every model. Categories with five to nine tasks
(Serialization, State Preparation, Visualization, Quantum Information) have
intervals spanning 40--80\,pp for mid-range models. Differences within these
categories should not be read as model rankings.

\section{QuantumBenchEval}
\label{app:qbe}

\subsection{Construction and Provenance}
\label{app:qbe-construction}

All 96 QBE tasks were written by the authors for this work and were not published or shared before evaluation. Problem settings use standard scientific instances (e.g., textbook Hamiltonians and small graph problems) rather than items from existing benchmarks. Each task specifies a problem statement, a required function signature, the return fields the candidate must produce, a reference solution, an assertion suite (T3 excepted), and a \texttt{common\_failure\_modes} field written by the task author.

\paragraph{Validation.} All 32 T2 and T6 reference solutions pass their assertion suites under both the original and corrected evaluators (Appendix~\ref{app:qbe-rescore}).

\paragraph{Known limitations.} Our audit found scoring contracts that check quantities the prompts do not specify (T1 and T6, corrected and rescored; T2, reported as recorded), three T2 tasks (\texttt{t2\_3}, \texttt{t2\_10}, \texttt{t2\_14}) whose disclosed data contradict their required answers, and assertions that are too permissive (Appendix~\ref{app:qbe-examples}). Task difficulty was not calibrated against human solvers, and topics are small (15--17 tasks), so per-topic intervals are wide.

\paragraph{Release.} QBE will be released with task files, reference solutions, assertion suites, the T3 rubric and judge prompt, and all saved outputs and scores under both the original and corrected evaluators.

\subsection{Topics and Scoring Contracts}
\label{app:qbe-topics}

QuantumBenchEval contains 96 tasks across six topics
(Table~\ref{tab:qbe-topics}). Each task specifies a required function signature
and a list of required output fields. Tasks in T1, T2, T4, T5, and T6 are scored
by executing the candidate and applying a task-specific assertion suite; T3 is
scored with a rubric (Appendix~\ref{app:t3}).

\begin{table}[ht]
\centering
\small
\setlength{\tabcolsep}{4pt}
\begin{tabular}{lp{4.6cm}cp{5.2cm}}
\toprule
Topic & Domain & Tasks & Scoring \\
\midrule
T1 & QAOA combinatorial optimization (e.g., weighted Max-Cut) & 17 & Assertions; corrected (undisclosed resource checks removed) \\
T2$^{*}$ & Constrained QUBO formulation & 15 & Assertions; undisclosed exact evaluation counts; 3 internally inconsistent tasks \\
T3 & VQE ground-state energy estimation & 15 & Execution + 10-point rubric \\
T4 & Many-body simulation and phase transitions & 15 & Assertions \\
T5 & Rydberg-blockade maximum independent set & 17 & Assertions \\
T6 & Graph partitioning & 17 & Assertions; corrected (undisclosed \texttt{reproducibility} form removed) \\
\bottomrule
\end{tabular}
\caption{QuantumBenchEval topics. T1 and T6 are rescored as described in Appendix~\ref{app:qbe-rescore}. $^{*}$T2 is reported as recorded; its contracts are corrected only in the released benchmark.}
\label{tab:qbe-topics}
\end{table}

\subsection{T3 Rubric and Judged Denominators}
\label{app:t3}

Each judged T3 sample is scored out of 10: energy accuracy (3 points), chemical
accuracy (1), required outputs (2), qubit and circuit specification (2), and
scientific reasoning (2). Candidate code runs in an isolated subprocess. Energy
accuracy is recomputed deterministically from the executed energy against the
task reference, and the remaining criteria are scored by
\texttt{claude-sonnet-4-6}. Truncated responses are neither executed nor judged
and are excluded rather than scored zero. Table~\ref{tab:t3} reports the
denominators.

\begin{table}[ht]
\centering
\small
\begin{tabular}{lcc}
\toprule
Model & Rubric mean & Judged / total \\
\midrule
GPT-6 Astra & 6.48 & 75/75 \\
Claude Fable 5.1$^{\S}$ & 7.77 & 75/75 \\
Gemini 3.1 Pro Preview & 4.87 & 75/75 \\
Gemini 3.6 Flash & 5.67 & 43/75 \\
Gemini 3 Flash & 4.99 & 75/75 \\
Gemma 3 4B & 1.59 & 75/75 \\
LLaMA 3.2 3B Instruct & 1.14 & 70/75 \\
Mistral 7B Instruct & 1.40 & 75/75 \\
Granite-Qiskit 8B & n/a & 0/75 \\
Qwen-Qiskit 14B & 1.99 & 67/75 \\
\bottomrule
\end{tabular}
\caption{T3 conditional rubric means. $^{\S}$Judge shares a provider with the evaluated model.}
\label{tab:t3}
\end{table}

\paragraph{Limitations of T3 scoring.} T3 scores should be read as indicative
rather than calibrated. (i)~Seven of the ten rubric points are assigned by a
single LLM judge (\texttt{claude-sonnet-4-6}) in one pass; we did not measure
inter-run agreement, agreement with a second judge, or agreement with human
raters. (ii)~The judge shares a provider with Claude Fable~5.1, the
highest-scoring model on T3 ($^{\S}$), so self-preference cannot be ruled out.
(iii)~Means are conditional on judged samples, and denominators differ across
models: Gemini 3.6 Flash is scored on 43 of 75 samples and Granite-Qiskit on none,
so its means are not comparable to models judged on all 75. (iv)~Only the
3-point energy-accuracy term is computed deterministically; the other criteria
(required outputs, circuit specification, scientific reasoning) depend on the
judge's reading of code and execution output. (v)~The rubric weights were set by
the authors and not validated against expert rankings. A human-rated subset and a
cross-provider judge would address (i)--(ii); we leave both to the released
benchmark.

\subsection{T2 and T6 Rescoring}
\label{app:qbe-rescore}

Before changing any evaluator, we audited every T2 and T6 assertion against its
candidate-facing prompt, then applied a pre-declared minimal change. For T6, only
the \texttt{reproducibility} assertion was removed, because no prompt defines its
form; all partition-quality assertions are unchanged. For T2, only exact
evaluation-count assertions were removed, and tasks \texttt{t2\_3},
\texttt{t2\_10}, and \texttt{t2\_14} were excluded as internally inconsistent
(their disclosed data contradict their required answers). No selection encoding
was normalized. Saved outputs were replayed without regeneration. All 32 reference
solutions pass under both evaluators, and no paired sample changes from pass to
fail. All 99 T6 flips come from samples whose first recorded failure was the
\texttt{reproducibility} type check. The replay ran on x86\_64, whereas the
original runs used aarch64; 5 of 1{,}600 samples differ between historical and
replayed original outcomes because of simulator stochasticity. One of these
(Gemini~3.1 Pro, \texttt{t6\_8.3}) fails an unchanged quality assertion under both
evaluators.

\begin{table}[ht]
\centering
\scriptsize
\begin{tabular}{lcccc}
\toprule
Model & T2 orig.\ (15) & T2 v2 (12) & T6 orig.\ (17) & T6 v2 (17) \\
\midrule
GPT-6 Astra & 4.0 [0.0--12.0] & 13.3 [0.0--33.3] & 1.2 [0.0--3.5] & 92.9 [80.0--100.0] \\
Claude Fable 5.1 & 1.3 [0.0--4.0] & 11.7 [0.0--30.0] & 70.6 [55.3--84.7] & 84.7 [75.3--92.9] \\
Gemini 3.1 Pro Preview & 0.0 & 8.3 [0.0--25.0] & 21.2 [14.1--28.2] & 27.1 [18.8--36.5] \\
Gemini 3.6 Flash & 2.7 [0.0--6.7] & 8.3 [0.0--20.0] & 76.5 [67.1--85.9] & 81.2 [74.1--88.2] \\
Gemini 3 Flash & 0.0 & 0.0 & 28.2 [17.6--41.2] & 31.8 [20.0--44.7] \\
Gemma 3 4B, LLaMA 3.2 3B, Mistral 7B, & 0.0 & 0.0 & 0.0 & 0.0 \\
\quad Granite-Qiskit 8B, Qwen-Qiskit 14B & & & & \\
\bottomrule
\end{tabular}
\caption{T2 and T6 pass@1 (\%) before and after rescoring, with 95\% task-bootstrap
intervals. T2 v2 scores a different task set (12 of 15 tasks), so part of that
change reflects the task population.}
\label{tab:qbe-rescore}
\end{table}

\subsection{Worked Examples}
\label{app:qbe-examples}

We give one task per topic. For each, we contrast a reference solution that
passes the assertion suite with a constructed failing response that reproduces a
failure mode observed during benchmark construction or anticipated by the task's
\texttt{common\_failure\_modes} field. These are illustrative responses, not
model outputs. Problem statements are summarized here; full statements and both
responses are in the supplementary material. Three of the six examples also
expose a gap in the task's own assertions (T1, T5, T6). We report these as
evaluator limitations of the same kind audited in Section~\ref{sec:qbe_results}.

\paragraph{T1 (\texttt{t1\_0}): weighted Max-Cut on a 4-cycle.} Partition four
substations joined in a ring (edge loads 4, 6, 3, 5\,MW) to maximize the
crossing load, using a Qiskit QAOA workflow. Required outputs include
\texttt{approximation\_ratio}, \texttt{success\_probability},
\texttt{circuit\_depth}, and \texttt{two\_qubit\_gate\_count}.
\emph{Failure:} The response returns the adjacent-pair partition, which cuts
$6{+}5{=}11$ rather than the bipartite optimum of 18, so
\texttt{cut\_value == 18} fails. It also counts two-qubit gates by name
(\texttt{cx}/\texttt{cz}/\texttt{ecr}), which returns 0 for a transpiled circuit
containing four \texttt{rzz} gates. \emph{Assertion gap:} The second error is
not caught, because the shipped check is \texttt{two\_qubit\_gate\_count >= 0}.

\paragraph{T2 (\texttt{t2\_0}): penalty-constrained job selection.} Select
exactly two of four jobs (priorities 4, 6, 5, 3) using an explicitly given QUBO
whose penalty term $A(\sum_i x_i-2)^2$ with $A{=}8$ enforces the cardinality
constraint. \emph{Failure:} The response optimizes the bare objective and ignores
the penalty, so it selects all four jobs (priority 18, infeasible) and reports
a constraint-violation rate of 0 that it never measured. The checks
\texttt{achieved\_priority == 11} and \texttt{selected\_jobs == \{2,3\}} both
fail; by the benchmark's own QUBO, its energy is $-18$ against $-43$ for
$\{J_2,J_3\}$. Both checked fields are part of the candidate-facing return
interface, so this is a genuine solution error rather than a disclosure issue.

\paragraph{T3 (\texttt{t3\_0}): VQE for H$_2$/STO-3G.} Estimate the
ground-state energy of the 2-qubit tapered H$_2$ Hamiltonian at equilibrium
(reference $E_\mathrm{FCI}{=}-1.13727$\,Ha; chemical accuracy $1.6$\,mHa).
\emph{Failure:} The response prepares the Hartree--Fock state $|01\rangle$ and
never varies it, reporting $-1.11671$\,Ha. This misses the full 20.6\,mHa of
correlation energy while asserting \texttt{chemical\_accuracy\_achievement =
True}. It also measures all six Pauli terms separately rather than the three
qubit-wise-commuting groups, which doubles \texttt{measurement\_cost}.

\paragraph{T4 (\texttt{t4\_0}): phase transition in a 4-site TFIM.} Sweep
$h/J\in[0.1,3.0]$ for an open transverse-field Ising chain and locate the
pseudo-critical field from the peak of the longitudinal susceptibility.
\emph{Failure:} Because $H$ commutes with $\prod_i\sigma^x_i$ and the
finite-chain ground state is non-degenerate, $\langle\sigma^z_i\rangle$ vanishes
identically (below $10^{-14}$ across the sweep). The computed susceptibility is
therefore differentiated floating-point noise, and its peak falls at the first
sweep point. The response returns $h_c{=}0.10$, failing
$0.30\le h_c\le 1.10$, and labels every point paramagnetic, failing
\texttt{phase\_classification\_accuracy > 0.70}. The reference solution returns
$h_c{=}0.60$.

\paragraph{T5 (\texttt{t5\_0}): maximum independent set on a Rydberg chain.}
Find the largest simultaneously excitable subset of five atoms whose blockade
graph is the path $P_5$ (optimum size 3). \emph{Failure:} A highest-degree-first
greedy rule selects $\{1,3\}$ and blocks the optimum $\{0,2,4\}$, so the
reported ratio of $0.667$ fails \texttt{independent\_set\_ratio >= 1 - 1e-6}.
\emph{Assertion gap:} The response computes \texttt{embedding\_overhead} on the
untranspiled circuit, giving exactly 1.0. This satisfies the shipped bound but
says nothing about SWAP inflation under the linear coupling map that the metric
is meant to capture.

\paragraph{T6 (\texttt{t6\_0}): NUMA-aware graph bisection.} Partition four
cores (two sockets, one cross-link) into two domains minimizing cross-domain
edges; the reference cut is 1. \emph{Failure:} Inverting the two Lagrange weights
lets balance dominate. The response returns $\{P_0,P_2\}\mid\{P_1,P_3\}$, which
is balanced but severs all three edges, so the approximation ratio of $0.333$
fails. \emph{Assertion gap:} A response with no balance term at all returns the
trivial one-sided assignment (cut 0). Because the task sets
\texttt{approximation\_ratio} to 1.0 whenever the achieved cut does not exceed
the reference, and \texttt{balance\_constraint\_score} $=0$ lies inside the
asserted $[0,1]$ range, this invalid non-bisection passes every check. The
problem statement explicitly forbids it, yet it is rated optimal.

\section{Additional Results}
\label{app:results}

\subsection{Full Pass@k Results}
\label{app:passk}

Table~\ref{tab:passk-full} reports pass@1, 3, and 5 for the four
execution-scored suites; QCoder is reported separately in
Table~\ref{tab:qcoder_full}.

\begin{table}[ht]
\centering
\scriptsize
\setlength{\tabcolsep}{2.5pt}
\resizebox{\linewidth}{!}{%
\begin{tabular}{l ccc ccc ccc ccc}
\toprule
& \multicolumn{3}{c}{QHE Standard} & \multicolumn{3}{c}{QHE Hard} & \multicolumn{3}{c}{QuanBench-44$^{\ddagger}$} & \multicolumn{3}{c}{QuanBench-117$^{\ddagger}$} \\
\cmidrule(lr){2-4}\cmidrule(lr){5-7}\cmidrule(lr){8-10}\cmidrule(lr){11-13}
Model & P@1 & P@3 & P@5 & P@1 & P@3 & P@5 & P@1 & P@3 & P@5 & P@1 & P@3 & P@5 \\
\midrule
GPT-6 Astra & 98.7 & 98.7 & 98.7 & 63.8 & 67.5 & 68.9 & 54.5 & 54.5 & 54.5 & 84.6 & 84.6 & 84.6 \\
Gemini 3.1 Pro Preview & 97.6 & 98.0 & 98.0 & 58.1 & 64.2 & 64.9 & 57.3 & 58.2 & 59.1 & 83.1 & 84.4 & 84.6 \\
Claude Fable 5.1 & 70.7 & 70.9 & 70.9 & 56.0 & 60.7 & 62.3 & 60.0 & 64.1 & 65.9 & 84.6 & 84.6 & 84.6 \\
Claude Opus 4.6 & 71.7 & 71.9 & 72.2 & 44.1 & 45.6 & 45.7 & 52.3 & 56.8 & 59.1 & 84.4 & 85.4 & 85.5 \\
Gemini 3.6 Flash & 64.0 & 68.3 & 69.5 & 40.8 & 46.9 & 48.3 & 31.8 & 41.1 & 45.5 & 50.6 & 65.6 & 71.8 \\
Gemini 3 Flash & 62.8 & 68.7 & 70.2 & 32.6 & 39.9 & 43.0 & 25.5 & 38.2 & 43.2 & 44.4 & 62.3 & 69.2 \\
GPT-5$^{\diamond}$ & 46.5 & 54.0 & 57.0 & 24.1 & 30.1 & 32.5 & 58.6 & 66.6 & 68.2 & 81.0 & 85.4 & 85.5 \\
\midrule
Muse-Glimmer-30B & 95.1 & 98.5 & 98.7 & 28.3 & 36.4 & 39.1 & 45.9 & 57.0 & 59.1 & 81.9 & 84.6 & 84.6 \\
Gemma-3-4B & 65.2 & 67.9 & 69.5 & 0.1 & 0.4 & 0.7 & 11.8 & 13.6 & 13.6 & 77.6 & 83.8 & 84.6 \\
LLaMA-3.1-8B & 5.7 & 14.0 & 20.5 & 0.1 & 0.4 & 0.7 & -- & -- & -- & 28.0 & 54.5 & 65.0 \\
Mistral-7B & 9.4 & 17.0 & 21.9 & 0.4 & 1.2 & 2.0 & 5.0 & 8.0 & 9.1 & 11.6 & 16.0 & 17.9 \\
\midrule
Mistral-Qiskit & 67.2 & 70.8 & 70.9 & 23.4 & 31.7 & 35.1 & 25.9 & 32.5 & 36.4 & 79.8 & 84.8 & 85.5 \\
Qwen-Qiskit & 53.8 & 70.3 & 72.2 & 22.3 & 32.3 & 35.1 & 27.7 & 37.0 & 40.9 & 74.5 & 82.4 & 83.8 \\
Granite-Qiskit & 20.5 & 36.5 & 47.7 & 11.4 & 20.0 & 24.5 & 10.5 & 25.5 & 36.4 & 31.3 & 59.4 & 69.2 \\
\midrule
Quantum RAG & 41.3 & 62.9 & 70.2 & 8.5 & 15.2 & 18.5 & -- & -- & -- & 42.1 & 58.0 & 61.5 \\
\bottomrule
\end{tabular}}
\caption{Pass@$k$ (\%) on the execution-scored suites. Per-cell 95\% intervals
are in the released artifacts. ``--'': the saved QuanBench-44 generations for
LLaMA-3.1-8B and Quantum RAG duplicate their QuanBench-117 outputs and were not
evaluated on QuanBench-44. $^{\ddagger}$Recorded evaluator outcomes pending
replay. $^{\diamond}$High empty-output rate
(Appendix~\ref{app:validity}).}
\label{tab:passk-full}
\end{table}

\subsection{Standard--Hard Gaps and Observed Coverage}
\label{app:coverage}

Table~\ref{tab:std-hard} gives the matched-task difference between Standard and
Hard pass@1 for each model. The gap ranges from 9\,pp (Granite-Qiskit) to
67\,pp (Muse-Glimmer-30B). Models with near-identical Standard scores diverge
sharply on Hard: Gemma-3-4B and Gemini 3.6 Flash both score about 65\% on
Standard but 0.1\% and 40.8\% on Hard. Because Hard changes the prompt
scaffolding (Appendix~\ref{app:examples}), these gaps reflect sensitivity to
format and scaffolding, not only task difficulty.

\begin{table}[ht]
\centering
\scriptsize
\begin{tabular}{lc@{\hspace{16pt}}lc}
\toprule
Model & Std$-$Hard, pp [95\% CI] & Model & Std$-$Hard, pp [95\% CI] \\
\midrule
GPT-6 Astra & 34.8 [27.4, 42.1] & Gemma-3-4B & 65.0 [57.6, 72.2] \\
Gemini 3.1 Pro Preview & 39.5 [32.1, 46.9] & LLaMA-3.1-8B & 5.6 [3.6, 7.8] \\
Claude Fable 5.1 & 14.7 [6.9, 22.5] & Mistral-7B & 9.0 [5.7, 12.6] \\
Claude Opus 4.6 & 27.6 [20.7, 34.6] & Mistral-Qiskit & 43.7 [36.6, 50.9] \\
Gemini 3.6 Flash & 23.2 [17.2, 29.3] & Qwen-Qiskit & 31.5 [25.2, 38.0] \\
Gemini 3 Flash & 30.2 [24.0, 36.6] & Granite-Qiskit & 9.1 [2.8, 15.8] \\
GPT-5$^{\diamond}$ & 22.4 [17.0, 28.1] & Quantum RAG & 32.9 [27.6, 38.4] \\
Muse-Glimmer-30B & 66.8 [60.3, 73.1] & & \\
\bottomrule
\end{tabular}
\caption{Matched Standard$-$Hard pass@1 differences (151 shared task IDs,
paired task bootstrap).}
\label{tab:std-hard}
\end{table}

\paragraph{Observed coverage.} Table~\ref{tab:coverage} counts the tasks that no
model solves in any of its five samples. These are observed-coverage statistics,
not evidence that a task is unsolvable. The 17 unsolved QuanBench-117 tasks share
a benchmark-side import obstruction (Section~\ref{sec:quanbench_audit}).

\begin{table}[ht]
\centering
\small
\begin{tabular}{lccc}
\toprule
Suite & Models & Tasks & Unsolved by all (5 samples) \\
\midrule
QHE Standard & 15 & 151 & 1 \\
QHE Hard & 15 & 151 & 31 \\
QuanBench-44$^{\ddagger}$ & 13 & 44 & 9 \\
QuanBench-117$^{\ddagger}$ & 15 & 117 & 17 \\
QCoder$^{\dagger}$ & 13 & 67 & 12 \\
\bottomrule
\end{tabular}
\caption{Tasks unsolved across the full roster. Task IDs are in the released
artifacts.}
\label{tab:coverage}
\end{table}

\begin{figure}[ht]
\centering
\includegraphics[width=\linewidth]{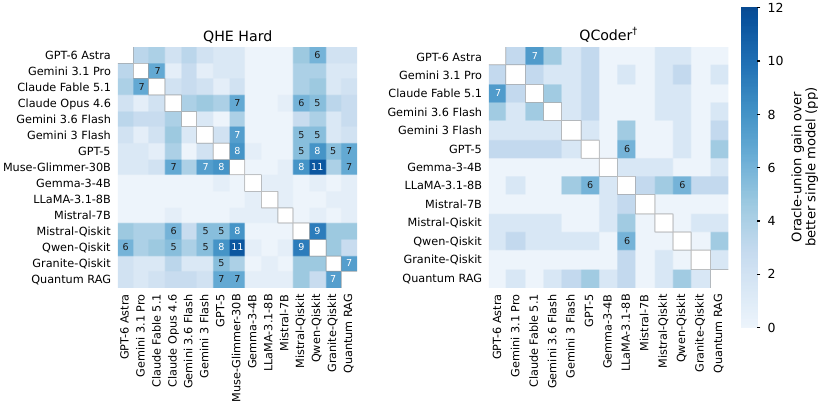}
\caption{Pairwise oracle-union gain in observed coverage (pp over the better
model of each pair). Values $\ge$5\,pp are labeled.}
\label{fig:complementarity}
\end{figure}

\paragraph{Complementarity.} Figure~\ref{fig:complementarity} shows, for each
model pair, how many more tasks the oracle union of their solved sets covers
than the better single model. Gains are modest (at most 11\,pp on Hard and 7\,pp
on QCoder). On Hard they are largest for pairs involving quantum-specialized models
(Muse-Glimmer-30B with Qwen-Qiskit, 11\,pp; Mistral-Qiskit with Qwen-Qiskit,
9\,pp), which suggests that specialized fine-tuning covers some tasks that
general models miss. On QCoder the largest gain is between the two strongest
models, Claude Fable~5.1 and GPT-6 Astra (7\,pp). Union
coverage is an oracle quantity with double the generation budget; it is not the
performance of a deployable ensemble.

\begin{figure}[ht]
\centering
\includegraphics[width=\linewidth]{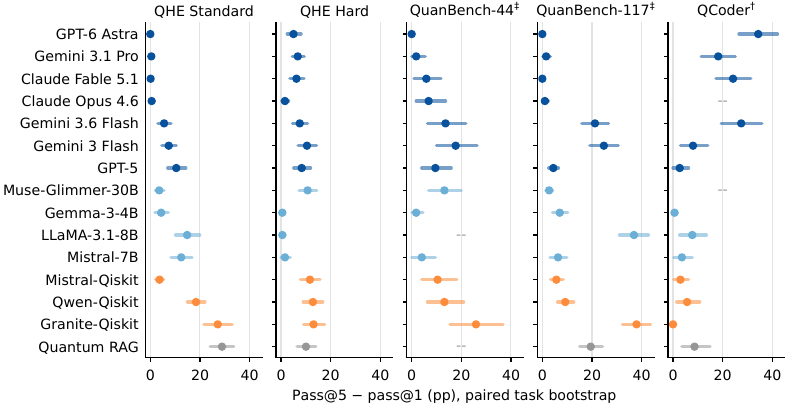}
\caption{Paired pass@5$-$pass@1 gain (pp) with 95\% task-bootstrap intervals.
$^{\dagger}$QCoder samples are prior-conditioned rather than i.i.d.; its gains
reflect the diversification protocol and are not comparable across suites.}
\label{fig:sampling-gain}
\end{figure}

\paragraph{Sampling gain.} Figure~\ref{fig:sampling-gain} shows the paired gain
from pass@1 to pass@5. The strongest hosted models gain almost nothing on the
execution-scored suites: GPT-6 Astra's gain is exactly zero on Standard,
QuanBench-44, and QuanBench-117, and Claude Fable~5.1's is zero on QuanBench-117
despite sampling at temperature 1.0. For these models, the five samples per task
almost always succeed or fail together, so pass@5 adds little beyond pass@1.
Quantum-specialized and small open-weight models gain far more, up to 38\,pp
(Granite-Qiskit on QuanBench-117), which reorders several models between pass@1
and pass@5 (Appendix~\ref{app:stats}). The Gemini Flash models also gain
21--25\,pp on QuanBench-117, the suite on which their syntax-valid rate is
lowest (Table~\ref{tab:validity}). This pattern is consistent with intermittent
formatting failures that additional samples recover, although we have not
verified that cause. QCoder gains are largest for the strongest models (up to
34\,pp for GPT-6 Astra). Because QCoder conditions each sample on the model's
prior attempts to diversify outputs, these gains are not comparable to those on
the i.i.d.\ suites.

\subsection{QCoder: Full Results and Judge Protocol}
\label{app:qcoder}

QCoder tasks have no canonical solutions or executable tests. Each candidate
therefore receives a binary \textsc{Pass}/\textsc{Fail} verdict from an LLM judge
given the problem statement and candidate code (Table~\ref{tab:judges}). All
QCoder judges are Anthropic models: \texttt{claude-sonnet-4-20250514} for runs
before April 2026 and \texttt{claude-sonnet-4-6} afterwards. Claude Opus~4.6 has
no QCoder run. Each sample is conditioned on the model's prior attempts to reduce
duplicates. Because this violates the i.i.d.\ assumption of the pass@$k$
estimator, the values below describe the recorded sequential protocol rather than
unbiased independent-sampling estimates.

Judge verdicts are not deterministic: identical code can receive different
verdicts in separate judgment calls (QCoder example in Appendix~\ref{app:qualitative}). We measured neither inter-run nor
cross-provider judge agreement, so QCoder results should be read as
judge-dependent.

\begin{table}[ht]
\centering
\small
\setlength{\tabcolsep}{4pt}
\begin{tabular}{lccc}
\toprule
Model & Pass@1 & Pass@3 & Pass@5 \\
\midrule
\multicolumn{4}{l}{\textit{Quantum-specialized}} \\
Granite-Qiskit & 7.5 [1.5--14.9] & 7.5 [1.5--14.9] & 7.5 [1.5--14.9] \\
Qwen-Qiskit & 4.8 [1.2--9.3] & 8.4 [2.8--14.8] & 10.4 [3.0--17.9] \\
Mistral-Qiskit & 4.5 [1.2--9.0] & 7.0 [1.3--13.9] & 7.5 [1.5--14.9] \\
\multicolumn{4}{l}{\textit{Retrieval-augmented}} \\
Quantum RAG & 9.3 [3.9--15.5] & 14.3 [7.2--22.4] & 17.9 [9.0--26.9] \\
\multicolumn{4}{l}{\textit{General-purpose open-weight}} \\
LLaMA-3.1-8B & 4.2 [1.2--8.1] & 8.4 [3.3--14.3] & 11.9 [4.5--20.9] \\
Mistral-7B & 2.4 [0.3--5.4] & 4.6 [0.9--9.7] & 6.0 [1.5--11.9] \\
Gemma-3-4B & 3.9 [0.0--8.7] & 4.5 [0.0--10.4] & 4.5 [0.0--10.4] \\
\multicolumn{4}{l}{\textit{Hosted}} \\
GPT-5$^{\diamond}$ & 7.8 [2.1--14.6] & 9.3 [3.0--16.4] & 10.4 [4.5--17.9] \\
Gemini 3 Flash & 2.4 [0.9--4.2] & 6.7 [2.2--12.1] & 10.4 [4.5--17.9] \\
Gemini 3.6 Flash & 23.3 [16.4--30.7] & 41.2 [30.9--51.3] & 50.7 [38.8--62.7] \\
Gemini 3.1 Pro Preview & 22.1 [14.6--30.1] & 34.8 [24.3--45.5] & 40.3 [28.4--52.2] \\
GPT-6 Astra & 32.8 [25.4--40.6] & 56.9 [46.7--66.7] & 67.2 [55.2--77.6] \\
Claude Fable 5.1$^{\S}$ & 43.0 [33.7--52.2] & 60.3 [49.7--70.7] & 67.2 [55.2--77.6] \\
\bottomrule
\end{tabular}
\caption{QCoder pass@$k$ (\%) with pointwise 95\% task-bootstrap intervals under
the recorded sequential judge-scored protocol. $^{\S}$Judge shares a provider
with the evaluated model. $^{\diamond}$High empty-output rate
(Appendix~\ref{app:validity}).}
\label{tab:qcoder_full}
\end{table}

\subsection{Output Validity and CodeBLEU}
\label{app:validity}

\paragraph{Empty and truncated outputs.} GPT-5 returns empty saved code for
316/755 QiskitHumanEval-Standard samples (41.9\%), all scored as failures by
pass@$k$. Of these, 315 stop at exactly 2{,}048 output tokens, which is
consistent with the output budget being exhausted before any code is emitted.
GPT-5's empty-output rate grows on the suites with longer tasks: 41.9\% on
Standard, 53.0\% on Hard, 55.0\% on QuanBench-44, 59.5\% on QuanBench-117, and
63.0\% on QCoder. Table~\ref{tab:validity} reports nonempty and syntax-valid
rates for every model. Low syntax validity is not confined to small models:
Gemini 3 Flash and Gemini 3.6 Flash produce syntactically valid code for only
53\% and 59\% of QuanBench-117 samples, and Gemini 3.1 Pro for only 58\% of
QCoder samples.

\begin{table}[ht]
\centering
\scriptsize
\setlength{\tabcolsep}{4pt}
\begin{tabular}{lccccc}
\toprule
Model & QHE Std & QHE Hard & QB-44 & QB-117 & QCoder \\
\midrule
GPT-6 Astra & 100 / 100 & 100 / 100 & 100 / 100 & 100 / 100 & 100 / 100 \\
Gemini 3.1 Pro Preview & 100 / 100 & 100 / 100 & 100 / 100 & 99 / 98 & 99 / 58 \\
Claude Fable 5.1 & 100 / 100 & 100 / 100 & 100 / 100 & 100 / 100 & 100 / 99 \\
Claude Opus 4.6 & 100 / 100 & 100 / 100 & 100 / 100 & 100 / 99 & -- \\
Gemini 3.6 Flash & 100 / 82 & 99 / 76 & 100 / 60 & 100 / 59 & 100 / 53 \\
Gemini 3 Flash & 100 / 79 & 100 / 76 & 100 / 59 & 100 / 53 & 100 / 42 \\
GPT-5$^{\diamond}$ & 58 / 55 & 47 / 42 & 45 / 38 & 41 / 35 & 37 / 36 \\
Muse-Glimmer-30B & 100 / 100 & 100 / 99 & 100 / 95 & 97 / 96 & -- \\
Gemma-3-4B & 100 / 99 & 100 / 96 & 100 / 97 & 100 / 96 & 100 / 96 \\
LLaMA-3.1-8B & 86 / 30 & 94 / 17 & --$^{\ast}$ & 84 / 26 & 70 / 44 \\
Mistral-7B & 100 / 92 & 100 / 86 & 100 / 90 & 100 / 87 & 100 / 93 \\
Mistral-Qiskit & 100 / 100 & 100 / 100 & 100 / 100 & 92 / 92 & 100 / 100 \\
Qwen-Qiskit & 94 / 92 & 100 / 99 & 96 / 95 & 96 / 95 & 99 / 99 \\
Granite-Qiskit & 98 / 76 & 100 / 99 & 93 / 64 & 96 / 70 & 100 / 93 \\
Quantum RAG & 100 / 98 & 100 / 96 & --$^{\ast}$ & 100 / 96 & 100 / 92 \\
\bottomrule
\end{tabular}
\caption{Output validity: nonempty / syntax-valid samples as a percentage of all
samples (755, 755, 220, 585, and 335 per model). ``--'': no run;
``--$^{\ast}$'': QuanBench-44 generations duplicate QuanBench-117 and were not
tested.}
\label{tab:validity}
\end{table}

\begin{table}[ht]
\centering
\scriptsize
\begin{tabular}{lcccc}
\toprule
Model & QHE Std & QHE Hard & QB-44 & QB-117 \\
\midrule
GPT-6 Astra & 0.316 & 0.372 & 0.294 & 0.301 \\
Gemini 3.1 Pro Preview & 0.342 & 0.424 & 0.327 & 0.312 \\
Claude Fable 5.1 & 0.369 & 0.409 & 0.330 & 0.362 \\
Claude Opus 4.6 & 0.344 & 0.388 & 0.304 & 0.333 \\
Gemini 3.6 Flash & 0.260 & 0.305 & 0.172 & 0.224 \\
Gemini 3 Flash & 0.222 & 0.284 & 0.185 & 0.221 \\
GPT-5$^{\diamond}$ & 0.132 & 0.147 & 0.102 & 0.103 \\
Muse-Glimmer-30B & 0.308 & 0.331 & 0.300 & 0.310 \\
Gemma-3-4B & 0.251 & 0.262 & 0.267 & 0.319 \\
LLaMA-3.1-8B & 0.150 & 0.138 & -- & 0.142 \\
Mistral-7B & 0.255 & 0.276 & 0.220 & 0.243 \\
Mistral-Qiskit & 0.320 & 0.374 & 0.304 & 0.303 \\
Qwen-Qiskit & 0.277 & 0.359 & 0.246 & 0.267 \\
Granite-Qiskit & 0.282 & 0.324 & 0.169 & 0.200 \\
Quantum RAG & 0.256 & 0.295 & -- & 0.236 \\
\bottomrule
\end{tabular}
\caption{Corpus-level CodeBLEU against canonical solutions (sample 0 per task;
empty generations scored as empty, which depresses GPT-5). ``--'': no verified
cache match.}
\label{tab:codebleu}
\end{table}

\paragraph{Syntax validity.} All 2{,}650 audited nonempty GPT-6 Astra outputs
parse as Python, yet a substantial fraction fail evaluation. Saved execution
fields and import diagnostics cannot yet serve as a validated
infrastructure-adjusted execution-success rate, because candidate, reference,
dependency, and evaluator failures have not been fully adjudicated.

\paragraph{CodeBLEU.} Across 11{,}325 QiskitHumanEval-Standard samples,
per-sample CodeBLEU has a point-biserial correlation of $r{=}0.280$
[0.222--0.332] with recorded pass/fail; on Hard, $r{=}0.487$ [0.420--0.550].
Reference-code similarity carries information about correctness, but it is not
equivalent to functional success or quantum-semantic correctness.

\section{Qualitative Examples}
\label{app:qualitative}

We show representative passing and failing samples drawn from the recorded runs
(Claude Opus~4.6 for QiskitHumanEval and QuanBench; GPT-5 for QCoder). Full
responses are in the supplementary material. QuanBench pass labels come from the
evaluator audited in Section~\ref{sec:quanbench_audit} and have not been
revalidated. We therefore checked the QuanBench samples below independently by simulation.

\paragraph{QiskitHumanEval Standard, API mismatch (\texttt{qiskitHumanEval/4}).}
Both responses follow the same logic, but the failing one imports
\texttt{UnitaryGate} from \texttt{qiskit.quantum\_info}, which does not export
it, so the candidate raises \texttt{ImportError} before any circuit is built. The
error is a plausible but nonexistent import path rather than a logic error.

\paragraph{Problem Statement}
\begin{verbatim}
from qiskit import QuantumCircuit
def create_unitary_from_matrix():
    """ Write the function that converts the matrix [[0, 0, 0, 1],
    [0, 0, 1, 0],[1, 0, 0, 0],
    [0, 1, 0, 0]] into a unitary gate and apply it to a Quantum Circuit. 
    
    Then return the circuit.
    """
\end{verbatim}

\paragraph{Passing Response (\cmark)}
\begin{lstlisting}[style=quantumcode]
from qiskit import QuantumCircuit
from qiskit.quantum_info import Operator

def create_unitary_from_matrix():
    matrix = [[0, 0, 0, 1],
              [0, 0, 1, 0],
              [1, 0, 0, 0],
              [0, 1, 0, 0]]
    
    qc = QuantumCircuit(2)
    unitary_gate = Operator(matrix)
    qc.unitary(unitary_gate, [0, 1], label='custom_unitary')
    return qc
\end{lstlisting}

\paragraph{Failing Response (\xmark)}
\begin{lstlisting}[style=quantumcode]
from qiskit import QuantumCircuit
from qiskit.quantum_info import UnitaryGate
import numpy as np

def create_unitary_from_matrix():
    matrix = np.array([[0, 0, 0, 1],
                       [0, 0, 1, 0],
                       [1, 0, 0, 0],
                       [0, 1, 0, 0]])
    
    gate = UnitaryGate(matrix)
    
    qc = QuantumCircuit(2)
    qc.append(gate, [0, 1])
    
    return qc
\end{lstlisting}

\paragraph{QiskitHumanEval Hard, wrong framework (\texttt{qiskitHumanEval/110}).}
Without the Standard skeleton's Qiskit import, the failing response writes a
syntactically valid implementation in Cirq, which the Qiskit harness cannot
evaluate. Its first lines:

\begin{lstlisting}[style=quantumcode]
import cirq
import numpy as np
from cirq import Circuit, CliffordGate
import random

def equivalent_clifford_circuit(circuit: cirq.Circuit, n: int) -> list:
    """
    ...
\end{lstlisting}

\paragraph{QuanBench-44, repeated self-revision (Task \texttt{24}, 4-qubit W state).}
Both samples restart construction mid-generation. The recorded passing sample
clears the circuit twice (\texttt{qc.data.clear()}) before settling on a cascade
of controlled-$R_y$ and CNOT gates; we verified by statevector simulation that
the returned circuit prepares the 4-qubit W state exactly (fidelity 1.0). The
failing sample abandons its first circuit for a multi-controlled construction
that produces an unequal superposition (probabilities $0.125, 0.125, 0.5, 0.25$;
fidelity 0.21).

\paragraph{QuanBench-117, runtime search over decompositions (Task \texttt{110}, CX from iSWAP).}
Both samples embed several candidate decompositions and check them with
\texttt{Operator.equiv} at runtime. \emph{The recorded pass is a false pass:}
the sample defines the target function twice, and the circuit returned by the
second definition is not equivalent to CNOT up to global phase (checked with
\texttt{Operator.equiv} against both control orientations). It is a concrete
instance of the evaluator defects in Section~\ref{sec:quanbench_audit}. The
failing sample is truncated mid-statement (an unterminated \texttt{try:} block),
consistent with the 2{,}048-token output limit, so it fails as a
\texttt{SyntaxError} rather than as an incorrect decomposition.

\paragraph{QCoder, inconsistent judge verdicts (\texttt{QPC004\_A2}).}
The two responses below are character-for-character identical and correct (three
alternating CNOTs implement SWAP), yet in separate judgment calls by
\texttt{claude-sonnet-4-20250514} one was judged \textsc{Pass} and the other
\textsc{Fail}.

\paragraph{Problem Statement}
\begin{verbatim}
## Problem Statement
Implement the operation of swapping two qubit states on a quantum circuit 

$\mathrm{qc}$ with $2$ qubits.

More precisely, implement the operation defined as
$$
\begin{equation}
a_0\ket{00} + \underline{a_1}\ket{10} + \underline{a_2}\ket{01} + 
a_3\ket{11}  \xrightarrow{\mathrm{qc}} a_0\ket{00} + 
\underline{a_2}\ket{10} + \underline{a_1}\ket{01} + a_3\ket{11} \nonumber
\end{equation},
$$
where $a_i$ represents arbitrary probability amplitudes.

## Constraints
- **You cannnot use Qiskit's SwapGate in this problem.**
- Global phase is ignored in judge.
- The submitted code must follow the specified format:

    from qiskit import QuantumCircuit

    def solve() -> QuantumCircuit:
        qc = QuantumCircuit(2)
        # Write your code here:

        return qc
\end{verbatim}

\paragraph{Passing Response (\cmark)}
\begin{lstlisting}[style=quantumcode]
from qiskit import QuantumCircuit

def solve() -> QuantumCircuit:
    qc = QuantumCircuit(2)
    # Write your code here:
    qc.cx(0, 1)
    qc.cx(1, 0)
    qc.cx(0, 1)

    return qc
\end{lstlisting}

\paragraph{Failing Response (\xmark)}
\begin{lstlisting}[style=quantumcode]
from qiskit import QuantumCircuit

def solve() -> QuantumCircuit:
    qc = QuantumCircuit(2)
    # Write your code here:
    qc.cx(0, 1)
    qc.cx(1, 0)
    qc.cx(0, 1)

    return qc
\end{lstlisting}

%% file: iclr2027_conference.bib
@misc{riofrío2025quantumcomputingautomotiveapplications,
      title={Quantum Computing for Automotive Applications}, 
      author={Carlos A. Riofrío and Johannes Klepsch and Jernej Rudi Finžgar and Florian Kiwit and Leonhard Hölscher and Marvin Erdmann and Lukas Müller and Chandan Kumar and Youssef Achari Berrada and Andre Luckow},
      year={2025},
      eprint={2409.14183},
      archivePrefix={arXiv},
      primaryClass={quant-ph},
      url={https://arxiv.org/abs/2409.14183}, 
}

@misc{vishwakarma2024qiskit,
  title={Qiskit humaneval: An evaluation benchmark for quantum code generative models},
  author={Vishwakarma, Sanjay and Harkins, Francis and Golecha, Siddharth and Bajpe, Vishal Sharathchandra and Dupuis, Nicolas and Buratti, Luca and Kremer, David and Faro, Ismael and Puri, Ruchir and Cruz-Benito, Juan},
  booktitle={2024 IEEE International Conference on Quantum Computing and Engineering (QCE)},
  volume={1},
  pages={1169--1176},
  year={2024},
  organization={IEEE}
}

@misc{dupuis2024qiskit,
  title={Qiskit code assistant: Training llms for generating quantum computing code},
  author={Dupuis, Nicolas and Buratti, Luca and Vishwakarma, Sanjay and Forrat, Aitana Viudes and Kremer, David and Faro, Ismael and Puri, Ruchir and Cruz-Benito, Juan},
  booktitle={2024 IEEE LLM Aided Design Workshop (LAD)},
  pages={4},
  year={2024},
  organization={IEEE}
}

@misc{basit2025qhackbench,
  title={Qhackbench: Benchmarking large language models for quantum code generation using pennylane hackathon challenges},
  author={Basit, Abdul and Shao, Minghao and Asif, Muhammad Haider and Innan, Nouhaila and Kashif, Muhammad and Marchisio, Alberto and Shafique, Muhammad},
  journal={arXiv preprint arXiv:2506.20008},
  year={2025}
}

@misc{Basitetal_12025,
  author = {Abdul Basit and Minghao Shao and M. Asif and Nouhaila Innan and Muhammad Kashif and Alberto Marchisio and Muhammad Shafique},
  title = {PennyCoder: Efficient Domain-Specific LLMs for PennyLane-Based Quantum Code Generation},
  year = {2025},
  journal = {arXiv.org},
}

@misc{yang2024qcircuitnet,
  title={QCircuitNet: A Large-Scale Hierarchical Dataset for Quantum Algorithm Design},
  author={Yang, Rui and Gu, Yuntian and Wang, Ziruo and Liang, Yitao and Li, Tongyang},
  journal={arXiv preprint arXiv:2410.07961},
  year={2024}
}

@misc{kheiri2025qspark,
  title={QSpark: Towards Reliable Qiskit Code Generation},
  author={Kheiri, Kiana and Aamir, Aamna and Miranskyy, Andriy and Ding, Chen},
  journal={arXiv preprint arXiv:2507.12642},
  year={2025}
}

@misc{dupuis2025quantum,
  title={Quantum Verifiable Rewards for Post-Training Qiskit Code Assistant},
  author={Dupuis, Nicolas and Tiwari, Adarsh and Mroueh, Youssef and Kremer, David and Faro, Ismael and Cruz-Benito, Juan},
  journal={arXiv preprint arXiv:2508.20907},
  year={2025}
}

@misc{basit2025pennylang,
  title={Pennylang: Pioneering llm-based quantum code generation with a novel pennylane-centric dataset},
  author={Basit, Abdul and Innan, Nouhaila and Asif, Muhammad Haider and Shao, Minghao and Kashif, Muhammad and Marchisio, Alberto and Shafique, Muhammad},
  journal={arXiv preprint arXiv:2503.02497},
  year={2025}
}

@misc{aragones2024c4q,
  title={C4Q: A Chatbot for Quantum},
  author={Aragon{\'e}s-Soria, Yaiza and Oriol, Manuel},
  booktitle={Proceedings of the 5th ACM/IEEE International Workshop on Quantum Software Engineering},
  pages={29--36},
  year={2024}
}

@misc{apak2024ketgpt,
  title={Ketgpt--dataset augmentation of quantum circuits using transformers},
  author={Apak, Boran and Bandic, Medina and Sarkar, Aritra and Feld, Sebastian},
  booktitle={International Conference on Computational Science},
  pages={235--251},
  year={2024},
  publisher={Springer}
}

@misc{jern2025agent,
  title={Agent-Q: Fine-Tuning Large Language Models for Quantum Circuit Generation and Optimization},
  author={Jern, Linus and Uotila, Valter and Yu, Cong and Zhao, Bo},
  booktitle={2025 IEEE International Conference on Quantum Computing and Engineering (QCE). IEEE},
  year={2025}
}

@misc{mikuriya2025qcoder,
  title={QCoder Benchmark: Bridging Language Generation and Quantum Hardware through Simulator-Based Feedback},
  author={Mikuriya, Taku and Ishigaki, Tatsuya and Kawarada, Masayuki and Minami, Shunya and Kadowaki, Tadashi and Suzuki, Yohichi and Naito, Soshun and Takada, Shunya and Kato, Takumi and Basseda, Tamotsu and others},
  booktitle={Proceedings of the 18th International Natural Language Generation Conference},
  pages={743--752},
  year={2025}
}

@misc{chen2021evaluatingHumaneval,
      title={Evaluating Large Language Models Trained on Code},
      author={Mark Chen and Jerry Tworek and Heewoo Jun and Qiming Yuan and Henrique Ponde de Oliveira Pinto and Jared Kaplan and Harri Edwards and Yuri Burda and Nicholas Joseph and Greg Brockman and Alex Ray and Raul Puri and Gretchen Krueger and Michael Petrov and Heidy Khlaaf and Girish Sastry and Pamela Mishkin and Brooke Chan and Scott Gray and Nick Ryder and Mikhail Pavlov and Alethea Power and Lukasz Kaiser and Mohammad Bavarian and Clemens Winter and Philippe Tillet and Felipe Petroski Such and Dave Cummings and Matthias Plappert and Fotios Chantzis and Elizabeth Barnes and Ariel Herbert-Voss and William Hebgen Guss and Alex Nichol and Alex Paino and Nikolas Tezak and Jie Tang and Igor Babuschkin and Suchir Balaji and Shantanu Jain and William Saunders and Christopher Hesse and Andrew N. Carr and Jan Leike and Josh Achiam and Vedant Misra and Evan Morikawa and Alec Radford and Matthew Knight and Miles Brundage and Mira Murati and Katie Mayer and Peter Welinder and Bob McGrew and Dario Amodei and Sam McCandlish and Ilya Sutskever and Wojciech Zaremba},
      year={2021},
      eprint={2107.03374},
      archivePrefix={arXiv},
      primaryClass={cs.LG}
}

@misc{ren2020codebleumethodautomaticevaluation,
      title={CodeBLEU: a Method for Automatic Evaluation of Code Synthesis}, 
      author={Shuo Ren and Daya Guo and Shuai Lu and Long Zhou and Shujie Liu and Duyu Tang and Neel Sundaresan and Ming Zhou and Ambrosio Blanco and Shuai Ma},
      year={2020},
      eprint={2009.10297},
      archivePrefix={arXiv},
      primaryClass={cs.SE},
      url={https://arxiv.org/abs/2009.10297}, 
}

@article{singh2025openai,
  title={Openai gpt-5 system card},
  author={Singh, Aaditya and Fry, Adam and Perelman, Adam and Tart, Adam and Ganesh, Adi and El-Kishky, Ahmed and McLaughlin, Aidan and Low, Aiden and Ostrow, AJ and Ananthram, Akhila and others},
  journal={arXiv preprint arXiv:2601.03267},
  year={2025}
}

@article{liu2023your,
  title={Is your code generated by chatgpt really correct? rigorous evaluation of large language models for code generation},
  author={Liu, Jiawei and Xia, Chunqiu Steven and Wang, Yuyao and Zhang, Lingming},
  journal={Advances in neural information processing systems},
  volume={36},
  pages={21558--21572},
  year={2023}
}

@article{jimenez2023swe,
  title={Swe-bench: Can language models resolve real-world github issues?, 2024},
  author={Jimenez, Carlos E and Yang, John and Wettig, Alexander and Yao, Shunyu and Pei, Kexin and Press, Ofir and Narasimhan, Karthik},
  journal={URL https://arxiv. org/abs/2310.06770},
  volume={7},
  year={2023}
}

@inproceedings{jain2025livecodebench,
  title={Livecodebench: Holistic and contamination free evaluation of large language models for code},
  author={Jain, Naman and Gu, Alex and Li, Wen-Ding and Yan, Fanjia and Zhang, Tianjun and Wang, Sida and Solar-Lezama, Armando and Sen, Koushik and Stoica, Ion},
  booktitle={International Conference on Learning Representations},
  volume={2025},
  pages={58791--58831},
  year={2025}
}

@article{balunovic2026matharena,
  title={Matharena: Evaluating llms on uncontaminated math competitions},
  author={Balunovic, Mislav and Dekoninck, Jasper and Petrov, Ivo and Jovanovi{\'c}, Nikola and Vechev, Martin},
  journal={Advances in Neural Information Processing Systems},
  volume={38},
  year={2026}
}

@article{zheng2023judging,
  title={Judging llm-as-a-judge with mt-bench and chatbot arena},
  author={Zheng, Lianmin and Chiang, Wei-Lin and Sheng, Ying and Zhuang, Siyuan and Wu, Zhanghao and Zhuang, Yonghao and Lin, Zi and Li, Zhuohan and Li, Dacheng and Xing, Eric and others},
  journal={Advances in neural information processing systems},
  volume={36},
  pages={46595--46623},
  year={2023}
}

@article{miller2024adding,
  title={Adding error bars to evals: A statistical approach to language model evaluations},
  author={Miller, Evan},
  journal={arXiv preprint arXiv:2411.00640},
  year={2024}
}

@inproceedings{guo2025quanbench,
  title={Quanbench: Benchmarking quantum code generation with large language models},
  author={Guo, Xiaoyu and Wang, Minggu and Zhao, Jianjun},
  booktitle={2025 40th IEEE/ACM International Conference on Automated Software Engineering (ASE)},
  pages={2657--2669},
  year={2025},
  organization={IEEE}
}

@article{grattafiori2024llama,
  title={The llama 3 herd of models},
  author={Grattafiori, Aaron and Dubey, Abhimanyu and Jauhri, Abhinav and Pandey, Abhinav and Kadian, Abhishek and Al-Dahle, Ahmad and Letman, Aiesha and Mathur, Akhil and Schelten, Alan and Vaughan, Alex and others},
  journal={arXiv preprint arXiv:2407.21783},
  year={2024}
}

@misc{jiang2023mistral7b,
      title={Mistral 7B}, 
      author={Albert Q. Jiang and Alexandre Sablayrolles and Arthur Mensch and Chris Bamford and Devendra Singh Chaplot and Diego de las Casas and Florian Bressand and Gianna Lengyel and Guillaume Lample and Lucile Saulnier and Lélio Renard Lavaud and Marie-Anne Lachaux and Pierre Stock and Teven Le Scao and Thibaut Lavril and Thomas Wang and Timothée Lacroix and William El Sayed},
      year={2023},
      eprint={2310.06825},
      archivePrefix={arXiv},
      primaryClass={cs.CL},
      url={https://arxiv.org/abs/2310.06825}, 
}

@misc{gemmateam2025gemma3technicalreport,
      title={Gemma 3 Technical Report}, 
      author={Gemma Team and Aishwarya Kamath and Johan Ferret and Shreya Pathak and Nino Vieillard and Ramona Merhej and Sarah Perrin and Tatiana Matejovicova and Alexandre Ramé and Morgane Rivière and Louis Rouillard and Thomas Mesnard and Geoffrey Cideron and Jean-bastien Grill and Sabela Ramos and Edouard Yvinec and Michelle Casbon and Etienne Pot and Ivo Penchev and Gaël Liu and Francesco Visin and Kathleen Kenealy and Lucas Beyer and Xiaohai Zhai and Anton Tsitsulin and Robert Busa-Fekete and Alex Feng and Noveen Sachdeva and Benjamin Coleman and Yi Gao and Basil Mustafa and Iain Barr and Emilio Parisotto and David Tian and Matan Eyal and Colin Cherry and Jan-Thorsten Peter and Danila Sinopalnikov and Surya Bhupatiraju and Rishabh Agarwal and Mehran Kazemi and Dan Malkin and Ravin Kumar and David Vilar and Idan Brusilovsky and Jiaming Luo and Andreas Steiner and Abe Friesen and Abhanshu Sharma and Abheesht Sharma and Adi Mayrav Gilady and Adrian Goedeckemeyer and Alaa Saade and Alex Feng and Alexander Kolesnikov and Alexei Bendebury and Alvin Abdagic and Amit Vadi and András György and André Susano Pinto and Anil Das and Ankur Bapna and Antoine Miech and Antoine Yang and Antonia Paterson and Ashish Shenoy and Ayan Chakrabarti and Bilal Piot and Bo Wu and Bobak Shahriari and Bryce Petrini and Charlie Chen and Charline Le Lan and Christopher A. Choquette-Choo and CJ Carey and Cormac Brick and Daniel Deutsch and Danielle Eisenbud and Dee Cattle and Derek Cheng and Dimitris Paparas and Divyashree Shivakumar Sreepathihalli and Doug Reid and Dustin Tran and Dustin Zelle and Eric Noland and Erwin Huizenga and Eugene Kharitonov and Frederick Liu and Gagik Amirkhanyan and Glenn Cameron and Hadi Hashemi and Hanna Klimczak-Plucińska and Harman Singh and Harsh Mehta and Harshal Tushar Lehri and Hussein Hazimeh and Ian Ballantyne and Idan Szpektor and Ivan Nardini and Jean Pouget-Abadie and Jetha Chan and Joe Stanton and John Wieting and Jonathan Lai and Jordi Orbay and Joseph Fernandez and Josh Newlan and Ju-yeong Ji and Jyotinder Singh and Kat Black and Kathy Yu and Kevin Hui and Kiran Vodrahalli and Klaus Greff and Linhai Qiu and Marcella Valentine and Marina Coelho and Marvin Ritter and Matt Hoffman and Matthew Watson and Mayank Chaturvedi and Michael Moynihan and Min Ma and Nabila Babar and Natasha Noy and Nathan Byrd and Nick Roy and Nikola Momchev and Nilay Chauhan and Noveen Sachdeva and Oskar Bunyan and Pankil Botarda and Paul Caron and Paul Kishan Rubenstein and Phil Culliton and Philipp Schmid and Pier Giuseppe Sessa and Pingmei Xu and Piotr Stanczyk and Pouya Tafti and Rakesh Shivanna and Renjie Wu and Renke Pan and Reza Rokni and Rob Willoughby and Rohith Vallu and Ryan Mullins and Sammy Jerome and Sara Smoot and Sertan Girgin and Shariq Iqbal and Shashir Reddy and Shruti Sheth and Siim Põder and Sijal Bhatnagar and Sindhu Raghuram Panyam and Sivan Eiger and Susan Zhang and Tianqi Liu and Trevor Yacovone and Tyler Liechty and Uday Kalra and Utku Evci and Vedant Misra and Vincent Roseberry and Vlad Feinberg and Vlad Kolesnikov and Woohyun Han and Woosuk Kwon and Xi Chen and Yinlam Chow and Yuvein Zhu and Zichuan Wei and Zoltan Egyed and Victor Cotruta and Minh Giang and Phoebe Kirk and Anand Rao and Kat Black and Nabila Babar and Jessica Lo and Erica Moreira and Luiz Gustavo Martins and Omar Sanseviero and Lucas Gonzalez and Zach Gleicher and Tris Warkentin and Vahab Mirrokni and Evan Senter and Eli Collins and Joelle Barral and Zoubin Ghahramani and Raia Hadsell and Yossi Matias and D. Sculley and Slav Petrov and Noah Fiedel and Noam Shazeer and Oriol Vinyals and Jeff Dean and Demis Hassabis and Koray Kavukcuoglu and Clement Farabet and Elena Buchatskaya and Jean-Baptiste Alayrac and Rohan Anil and Dmitry and Lepikhin and Sebastian Borgeaud and Olivier Bachem and Armand Joulin and Alek Andreev and Cassidy Hardin and Robert Dadashi and Léonard Hussenot},
      year={2025},
      eprint={2503.19786},
      archivePrefix={arXiv},
      primaryClass={cs.CL},
      url={https://arxiv.org/abs/2503.19786}, 
}

@misc{lu2026museunifiedagenticharness,
      title={MUSE: A Unified Agentic Harness for MLLMs}, 
      author={Jianglin Lu and Hailing Wang and Xu Ma and Qihua Dong and Mingyuan Zhang and Yizhou Wang and Yun Fu},
      year={2026},
      eprint={2606.03005},
      archivePrefix={arXiv},
      primaryClass={cs.CV},
      url={https://arxiv.org/abs/2606.03005}, 
}

@INPROCEEDINGS{11544968,
  author={Stapleton, Nicholas and Afrose, Sharmin and Seefried, Ethan and Ortega, Vicente Leyton and Humble, Travis S. and Ghosal, Tirthankar},
  booktitle={2026 IEEE 19th Dallas Circuits and Systems Conference (DCAS)}, 
  title={Quantum-Rag: Retrieval-Augmented Generation Framework for Quantum Circuit Code Generation}, 
  year={2026},
  volume={},
  number={},
  pages={1-5},
  doi={10.1109/DCAS69364.2026.11544968}}

@misc{openai2026astra,
  author       = {{OpenAI}},
  title        = {{GPT-6 Astra}},
  year         = {2026},
  howpublished = {OpenAI API Documentation},
  note         = {Accessed 2026}
}

@misc{anthropic2026opus46,
  author       = {{Anthropic}},
  title        = {{Claude Opus 4.6}},
  year         = {2026},
  howpublished = {Claude Platform Documentation},
  note         = {Released February 5, 2026}
}

@misc{anthropic2026fable51,
  author       = {{Anthropic}},
  title        = {{Claude Fable 5.1}},
  year         = {2026},
  howpublished = {Claude Platform Documentation},
  note         = {Released September 1, 2026}
}

@misc{google2025gemini3flash,
  author       = {{Google DeepMind}},
  title        = {{Gemini 3 Flash}},
  year         = {2025},
  howpublished = {Gemini Model Documentation}
}

@misc{google2026gemini36flash,
  author       = {{Google}},
  title        = {{Gemini 3.6 Flash}},
  year         = {2026},
  howpublished = {Google AI},
  note         = {Released July 21, 2026}
}

@misc{google2026gemini31pro,
  author       = {{Google DeepMind}},
  title        = {{Gemini 3.1 Pro}},
  year         = {2026},
  howpublished = {Gemini Model Card},
  note         = {Published February 19, 2026}
}

@misc{qiskit2025qwen,
  author       = {{IBM Quantum}},
  title        = {{Qwen2.5-Coder-14B-Qiskit}},
  year         = {2025},
  howpublished = {Hugging Face Model Card}
}

@misc{qiskit2025mistral,
  author       = {{IBM Quantum}},
  title        = {{Mistral-Small-3.2-24B-Qiskit}},
  year         = {2025},
  howpublished = {Hugging Face Model Card}
}

@misc{qiskit2025granite,
  author       = {{IBM Quantum}},
  title        = {{Granite-3.2-8B-Qiskit}},
  year         = {2025},
  howpublished = {Hugging Face Model Card},
  note         = {Released June 3, 2025}
}
